\documentclass[12pt]{article}
\newcommand{\be}{\begin{equation}}
\newcommand{\ee}{\end{equation}}
\newcommand{\bea}{\begin{eqnarray}}
\newcommand{\eea}{\end{eqnarray}}

\usepackage{type1cm,amsmath,amssymb,color,graphics,amscd,amsfonts,mathrsfs,epsf,indentfirst}
\usepackage{epsfig}
\usepackage{bm}
\usepackage{subfigure}
\usepackage{multirow,graphicx}

\title{Black hole instabilities and local Penrose inequalities}

\author{Pau Figueras, Keiju Murata and Harvey S. Reall \\ {\small DAMTP, Centre for Mathematical Sciences, Wilberforce Road, Cambridge CB3 0WA, UK}}

\begin{document}

\maketitle

\begin{abstract}
Various higher-dimensional black holes have been shown to be unstable by studying linearized gravitational perturbations. A simpler method for demonstrating instability is to find initial data that describes a small perturbation of the black hole and violates a Penrose inequality. An easy way to construct initial data is by conformal rescaling of the unperturbed black hole initial data. For a compactified black string, we construct initial data which violates the inequality almost exactly where the Gregory-Laflamme instability appears. We then use the method to confirm the existence of the "ultraspinning" instability of Myers-Perry black holes. Finally we study black rings. We show that "fat" black rings are unstable. We find no evidence of any rotationally symmetric instability of "thin" black rings. \end{abstract}

\section{Introduction}

Vacuum black holes in four spacetime dimensions are believed to be stable against gravitational perturbations. A qualitatively new feature that emerges in higher dimensions is the possibility of black objects with unstable horizons. The first example to be discovered was the Gregory-Laflamme instability of a black string \cite{Gregory:1993vy}. Later, heuristic arguments were presented which suggest that  "ultraspinning" Myers-Perry \cite{Myers:1986un}
 black holes should suffer from a similar kind of instability  \cite{Emparan:2003sy}. The existence of this instability has been confirmed by studies of linearized perturbations \cite{Dias:2009iu,Dias:2010eu,Dias:2010maa,Dias:2011jg}.

The equations governing linearized perturbations of higher-dimensional rotating black holes are very complicated. It would be nice if there were a simpler method of demonstrating black hole instabilities. In this paper, we will show that the existence of certain types of instability can be demonstrated using inequalities analogous to the Penrose inequality (see Ref. \cite{Mars:2009cj} for a review). In the form presented in Ref. \cite{jang}, the 4d Penrose inequality is
\be
\label{penrose1}
 A_{\rm min} \le 16 \pi E^2
\ee
where $E$ is the ADM energy of asymptotically flat initial data for Einstein's equation and $A_{\rm min}$ is the greatest lower bound on the area of any surface that encloses the apparent horizon of this data ($A_{\rm min}$ may be less than the area of the apparent horizon \cite{horowitz}). This inequality results from the assumption that the spacetime resulting from initial data containing an apparent horizon must "settle down" to the Kerr solution. It has been proved only for the special case of time-symmetric initial data \cite{huisken,bray}. 

The inequality (\ref{penrose1}) is supposed to apply to any asymptotically flat initial data containing an apparent horizon. One can also consider a less general {\it local} Penrose inequality, in which (\ref{penrose1}) is restricted to initial data describing a small perturbation of a Schwarzschild black hole. Such an inequality was first investigated by Gibbons \cite{gibbons}, who considered certain time-symmetric initial data describing a small perturbation of the Schwarzschild solution. He found that (\ref{penrose1}) held to second order in perturbation theory for his data (at first order, the result follows from the first law of black hole mechanics \cite{Sudarsky:1992ty}). Related results were obtained in Ref. \cite{Hawking:1973aq}.

In this paper, we will use local Penrose inequalities to demonstrate instability of certain stationary black hole solutions. If such a black hole is stable, then initial data describing a small perturbation of the black hole must satisfy a local Penrose inequality analogous to (\ref{penrose1}). Therefore, if we can find initial data that violates the inequality then the black hole must be unstable. 

Our first example of this method is the Gregory-Laflamme instability. Consider a Schwarzschild black string compactified on a Kaluza-Klein circle of radius $2\pi L$. A surface of constant $t$ in this geometry gives initial data for Einstein's equation. Now consider new initial data corresponding to a small (but finite) perturbation of the black string (preserving the radius of the KK circle at infinity). Assume that the black string is stable. Then the perturbation will disperse through radiation to infinity and across the horizon so the evolution of this data will settle down to a new stationary black string. 
In general, this might have a small angular momentum $J_F$ and a small linear momentum $P_F$. The solution will be a boosted Kerr black string with mass $M_F$ and horizon area
\be
 A_F \le \frac{8 M_F^2}{L}
\ee
The inequality follows from the fact that a boosted Kerr string has smaller horizon area than a Schwarzschild black string of the same mass. Let $A_I$ denote the area of the intersection of the event horizon with the initial surface. Since the event horizon encloses the apparent horizon we have $A_{\rm min} \le A_I \le A_F$ where $A_{\rm min}$ was defined above and the second inequality follows from the second law. Gravitational waves carry away energy (the Bondi energy decreases) so $M_F \le E$  where $E$ is the ADM energy of the initial data. Combining these inequalities we are led to 
\be
\label{stringpenrose1}
 A_{\rm min} \le \frac{8 E^2}{L}
\ee
The inequality (\ref{stringpenrose1}) is a local Penrose inequality for the Schwarzschild black string. 

The assumption made in deriving (\ref{stringpenrose1}) is that the string is stable. Hence if we can find suitable initial data which violates the inequality then we have demonstrated that the string cannot be stable. We shall construct initial data describing a small perturbation of the black string simply by conformal rescaling of the initial data for the unperturbed string. Our initial data violates the local Penrose inequality when $r_+/L$ is smaller than a certain critical value (where $r_+$ is the horizon radius of the unperturbed string). The critical value for $r_+/L$ is smaller than the critical value at which the GL instability is known to appear by less than $0.2\%$. Hence the existence of the Gregory-Laflamme instability can be predicted from initial data alone. The close agreement between our result and the GL result is somewhat surprising: violation of the local Penrose inequality is a sufficient, but not necessary, condition for instability. 

Note that initial data corresponding to a constant $t$ surface in the black string spacetime saturates the inequality (\ref{stringpenrose1}). Hence, if the string is stable, such initial data minimizes $E$ for fixed $A_{\rm min}$ or maximizes $A_{\rm min}$ for fixed $E$. In other words, a stable string must be a local minimum of energy for fixed horizon area or a local maximum of horizon area for fixed energy in the space of asymptotically KK initial data for Einstein's equation. Our proof of instability amounts to showing that certain black strings fail to satisfy these properties.\footnote{A similar argument was used in Ref. \cite{Sudarsky:1992ty} to explain why coloured black hole solutions of Einstein-Yang-Mills theory should be unstable.}

The main aim of this paper is to use the above argument to demonstrate instabilities of higher-dimensional rotating black holes. In the rotating case, one needs to assume that the initial data preserves some rotational symmetry in order to derive a useful Penrose inequality \cite{gibbons}. Therefore this method can be used to demonstrate instabilities that preserve some rotational symmetry. This includes the Myers-Perry instabilities discovered in 
Refs.  \cite{Dias:2009iu,Dias:2010eu,Dias:2010maa,Dias:2011jg}. In this case, we shall confirm the existence of the "ultraspinning" instability predicted in Ref. \cite{Emparan:2003sy} with far less effort than required for the numerical analyses of Refs.  \cite{Dias:2009iu,Dias:2010maa}. However we cannot predict instabilities which break the relevant rotational symmetry, such as the "bar-mode" Myers-Perry instabilities found in Refs. \cite{Shibata:2009ad,Shibata:2010wz}. 

Next we consider the stability of the black ring solution of Ref. \cite{Emparan:2001wn}. So far, investigations of black ring stability have been heuristic. For a given mass, there is a finite range of angular momenta for which there exist two distinct ring solutions, referred to as "thin" and "fat" because of the shape of the horizon. Heuristic arguments suggest that "fat" rings should be unstable, as we now explain.

Ref. \cite{Arcioni:2004ww} used "turning point" methods to argue that fat rings must have one more "unstable mode" than thin rings. This argument assumes that the "states or configurations of a given system" correspond to points in some manifold ${\cal M}$ on which one can define quantities such as the mass, angular momenta, and entropy. "Equilibrium states" correspond to points which extremize the entropy at fixed mass and angular momenta. These are assumed to form a submanifold ${\cal M}_{\rm eq}$ of ${\cal M}$. In the present case, this corresponds to the known black ring solutions, hence one knows the entropy, etc., on ${\cal M}_{\rm eq}$. A given equilibrium state is stable if it is a local {\it maximum} of entropy for fixed mass and angular momentum. Stability can change at either a "turning point" or a "bifurcation". The former can be identified from knowledge of the thermodyamic quantities on ${\cal M}_{\rm eq}$. For black rings, there is a turning point as one moves from the thin ring to the fat ring branch. From this, one can deduce that fat rings near to the turning point are unstable. 

This argument is very suggestive (and successfully used in astrophysics to predict neutron star instabilities) but not entirely rigorous since, for example, the manifold ${\cal M}$ is not defined. A related point is that it is not clear whether the predicted instability should be present classically. Finally, it only allows one to deduce that fat rings are unstable near the turning point. It cannot be concluded that all fat rings are unstable because it is possible that there is another change in stability (at a bifurcation) as one moves along the fat ring branch. 

A different approach was taken in Ref. \cite{Elvang:2006dd}, which  considered certain singular deformations of the black ring to deduce an "effective potential" for variations of the radius of the ring. It was found that fat rings sit at a maximum of the potential and therefore should be unstable. Again, this is very suggestive (and in agreement with the result of Ref. \cite{Arcioni:2004ww}) but it is not clear that such a simple mechanical picture captures all of the relevant gravitational dynamics.

We will use violation of a local Penrose inequality to show that all fat black rings are indeed classically unstable. We will also construct multi-parameter families of initial data describing perturbations of thin black rings. These all respect the local Penrose inequality and so our results are consistent with thin rings being stable against rotationally symmetric perturbations. However, it is believed that thin rings with large enough angular momentum will suffer from a GL-like instability \cite{Emparan:2001wn}. Since this would involve breaking the rotational symmetry of the ring we cannot investigate it using our methods. 

Finally, we consider the "doubly spinning" black rings of Ref. \cite{Pomeransky:2006bd}. These also can be classified as "thin" and "fat". Presumably the arguments of Refs. \cite{Arcioni:2004ww,Elvang:2006dd} could also be applied to these but this has not been done so our work is the first study of the stability of doubly spinning rings. As in the singly spinning case, we find that fat black rings are unstable and we find no evidence of a rotationally-symmetric instability of thin rings.

\subsection*{Notation}

Initial data for the vacuum Einstein equation in $d$ dimensions will be denoted $(\Sigma,h_{ab},K_{ab})$ where $\Sigma$ is a $d-1$ manifold with Riemannian metric $h_{ab}$ and extrinsic curvature $K_{ab}$. We shall denote the apparent horizon on the initial data surface by $S$. $\nabla_a$ denotes the connection associated to $h_{ab}$. 
When discussing stationary black hole or black string solutions, $\Sigma$ will denote a surface of constant $t$ (in coordinates adapted to the timelike Killing field), which passes through the bifurcation surface where the past and future event horizons intersect. We will often consider 1-parameter families of initial data. The parameter will be denoted $\lambda$ and a derivative with respect to $\lambda$ denoted by a dot. For example, $\dot{h}_{ab}$ denotes a linearized perturbation of the metric of the initial data. An overbar denotes a quantity defined with respect to the unperturbed solution. We will work in units of $G=c=1$.

\section{Black string instability}
\label{sec:string}

Consider the black string in $d=n+3>4$ dimensions in the standard Schwarzschild coordinates:
\begin{equation}
ds^2=-f(r)\,dt^2+\frac{dr^2}{f(r)}+r^2\,d\Omega_{(n)}^2+dx^2\,,\quad f(r)=1-\frac{r_+^{n-1}}{r^{n-1}}\,,\quad x\sim x+2\pi\,L\,,
\end{equation}
where $r=r_+$ denotes the location of the horizon. Consider a $t=\textrm{constant}$ surface in this spacetime. To extend this surface to an Einstein-Rosen bridge with two asymptotically flat regions we define a new radial coordinate $y$ such that
\begin{equation}
r=\frac{r_+}{1-y^2}\,,\qquad 0\leq y< 1\,.
\end{equation}
In terms of this new coordinate the metric on our surface is
\begin{equation}
ds^2=\frac{4\,r_+^2\,dy^2}{g(y)(1-y^2)^4}+\frac{r_+^2}{(1-y^2)^2}\,d\Omega_{(n)}^2+dx^2\,,\quad g(y)=\frac{1-\left(1-y^2\right)^{n-1}}{y^2}\,.
\label{eqn:SchwBS}
\end{equation}
We can now analytically continue $y$ so that $y\in (-1,1)$. In these new coordinates the bifurcation surface is at $y=0$ and the two asymptotically flat regions are $y\to +1$ and $y\to -1$ respectively. $y \rightarrow -y$ is an isometry which interchanges these regions. We shall denote this surface by $\Sigma$, with metric $\bar{h}_{ab}$.  The extrinsic curvature of $\Sigma$ vanishes so $(\Sigma,\bar{h}_{ab})$ provides time-symmetric initial data for the vacuum Einstein equation.

We construct new time-symmetric initial data $(\Sigma,h_{ab})$ by the well-known method of conformal rescaling. Let
\be
 h_{ab} = \Psi^{\frac{4}{n}} \,\bar{h}_{ab}
\ee
The Hamiltonian constraint reduces to Laplace's equation:
\be
 \bar{\nabla}^2\Psi=0
 \ee
where $\bar{\nabla}^2$ is the Laplacian defined using $\bar{h}_{ab}$, and the momentum constraint is automatically satisfied. We seek solutions in which we excite just the lowest harmonic around the KK circle:
\be
\label{Psisol}
 \Psi = 1+ \lambda f(y) \cos( x/L)
\ee
where $\lambda$ is a parameter. The Laplace equation reduces to
\be
\label{feq}
f''(y)+\frac{(1-y^2)g'(y)+4(n-2)y\,g(y)}{2\,g(y)(1-y^2)}\,f'(y)-\frac{4\,r_+^2}{L^2\,g(y)(1-y^2)^4}\,f(y)= 0\,.
\ee
Solutions of this equation behave as $e^{\pm  r_+/(L(1-y^2))}$ as $y\rightarrow 1$ and are smooth at $y=0$. We fix $f(y)$ at some $y=y_\textrm{min}<0$ on the "other" side of the Einstein-Rosen bridge\footnote{To obtain the data presented in this section we chose $y_\textrm{min}=-0.1$. We have checked that changing the location of this inner surface does not alter our results. }  so that $f(y_\textrm{min})=1$, and we choose $f(y)$ to be the solution that decays as $e^{-r_+/(L(1-y^2))}$ as $y \rightarrow 1$.  We determine the solution numerically. Using this procedure we have constructed a 1-parameter family of initial data $(\Sigma,h_{ab})$ which reduces to the Schwarzschild black string initial data when $\lambda=0$. Note that $(\Sigma,h_{ab})$ has the same asymptotic behaviour as $(\Sigma,\bar{h}_{ab})$ as $y \rightarrow 1$. They differ in the other asymptotic region but this lies behind the horizon so this is not a problem. For small $\lambda$, our initial data describes a small perturbation of the black string.

Now we consider the local Penrose inequality (\ref{stringpenrose1}). If the string is stable then our initial data must satisfy this inequality for sufficiently small $\lambda$. Since our initial data is time-symmetric, the apparent horizon is an outermost minimal surface, i.e., it locally extremizes the area. Ref. \cite{Cai:2001su} showed that the minimal surface is {\it stable}, i.e., the extremum is a local minimum.\footnote{This point is discussed in more detail in Appendix \ref{sec:apparent}.} Therefore it has area no greater than any surface that encloses it and hence $A_{\rm min} = A_{\rm app}$, the area of the apparent horizon. Furthermore, time-symmetry implies that  the ADM energy is the same as the ADM mass, and so (\ref{stringpenrose1}) reduces to
\be
\label{stringpenrose2}
 A_{\rm app} \le \frac{8 M^2}{L}
\ee
where $M$ is the ADM mass of the initial data. Since $f(y)$ decays exponentially, it follows that $M$ is the same as the ADM mass of the unperturbed black string (i.e. $M$ does not depend on $\lambda$). The RHS of (\ref{stringpenrose2}) is just the area of the horizon of this string and so we can rewrite this as
$A_{\rm app}(\lambda) \le A_{\rm app}(0)$ for sufficiently small $\lambda$. Therefore we can prove that the string is unstable by showing that $A_{\rm app}(\lambda)>A_{\rm app}(0)$ for arbitrarily small $\lambda$. 

Since we are interested only in small $\lambda$, we can expand $A_{\rm app} = A_{\rm app}(0) +  \lambda \dot{A}_{\rm app} (0) + (1/2) \lambda^2 \ddot{A}_{\rm app} (0) + \ldots$ where a dot denotes a derivative with respect to $\lambda$. The first law of black hole mechanics applies to arbitrary linear perturbations that preserve the constraints on $\Sigma$ \cite{Sudarsky:1992ty}. The first law gives $\dot{A}(0) \propto \dot{M}(0)$. Here $A$ can denote the area of either the event horizon, or apparent horizon, since they agree to linear order.\footnote{
To see this, note that two effects might contribute to the change in $A$ at linear order. One is the change in the metric at $y=0$. The other is the change in the position of the horizon. But since $y=0$ is a minimal surface, the latter effect appears only at second order. Hence one has just the first effect, which does not depend on which kind of horizon we are discussing.}  For our initial data, $\dot{M}(0)=0$ and hence $\dot{A}_{\rm app}(0) =0$. Therefore our condition for instability reduces to
\be
 \ddot{A}_{\rm app}(0) > 0.
\ee
In Apppendix \ref{sec:horizonarea}, we explain how to calculate $\ddot{A}_{\rm app}(0)$. The result is:
\be
\label{d2Astring}
 \ddot{A}_{\rm app}(0) = \frac{(n+1)(n+2)}{n^2} \,A_{\rm app}(0) \left[ f(0)^2 - \frac{n^2-1}{(n+2)\big(n(n-1)+2\,r_+^2/L^2\big)}  \left( \frac{df}{dy} \right)_{y=0}^2 \right]
\ee
Hence we deduce that the black string is unstable if there exists a solution of the ODE (\ref{feq}) that decays as $y \rightarrow 1$ and satisfies
\be
   {\cal A} \equiv \left( \frac{1}{f} \frac{df}{dy} \right)_{y=0}^2 - \frac{n+2}{n^2-1} \left( n(n-1) + \frac{2 \,r_+^2}{L^2} \right) < 0
\ee

\begin{figure}[t]
\begin{center}
\includegraphics[scale=1]{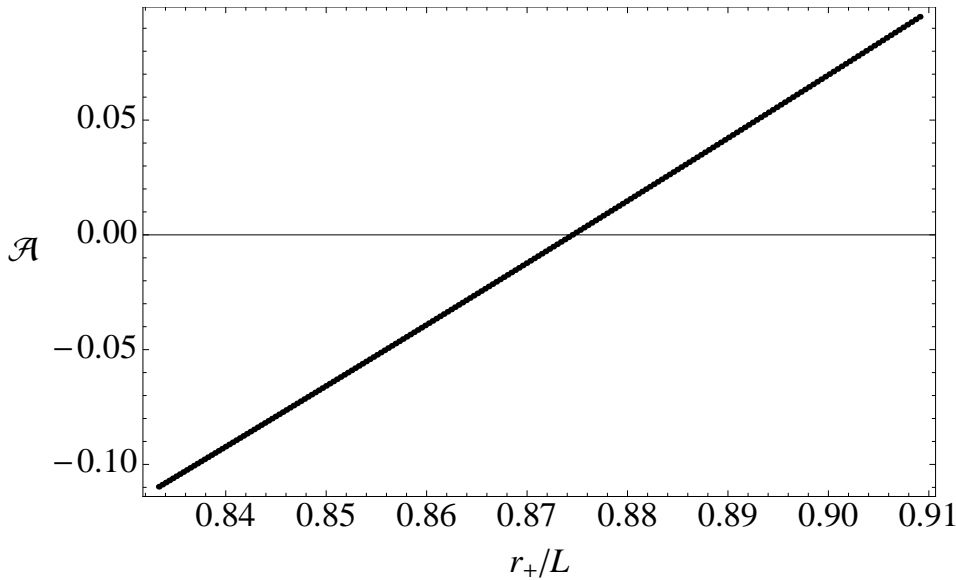}
\end{center}
\caption{${\cal A}$ vs. $r_+/L$  for the $d=5$ black string. For other dimensions the plots look qualitatively similar. For large values of $r_+/L$,  $\mathcal A$ is positive but it becomes negative at a certain critical value,  signalling an instability. For any number of dimensions this critical value is always smaller from critical value for the onset of the GL instability. In $d=5$ we find $(r_+/L)_\textrm{crit}=0.8745$, which differs by less than $0.2$\% from the GL critical value. The agreement gets better as the number of dimensions increases.  }
\label{fig:BS}
\end{figure}

We have solved (\ref{feq}) numerically and Figure \ref{fig:BS} shows ${\cal A}$ as a function of $r_+/L$ for the $d=5$ case. Note that ${\cal A}$ is positive for large $r_+/L$, consistent with stability, but becomes negative at small $r_+/L$, indicating instability. In terms of $r_+$, the critical value is $r_+/L = 0.8745$. This can be compared with the critical value below which the GL instability exists:\footnote{
We could not find a result to this accuracy in the literature so we determined it ourselves. The same applies for the other values quoted in Table \ref{table:BS}.} $r_+/L = 0.8762$. The two results agree to an accuracy of $0.2 \%$.

We have performed an analogous calculation for Schwarzschild black strings with $d=6,\ldots,\,11$. The results are shown in Table \ref{table:BS}. Our value is always smaller than the GL value and the agreement between the two values becomes better as $d$ increases. 
\begin{table}[h]
\begin{center}
\begin{tabular}{c|c|c|c|c|c|c|c}
$d$ & 5 & 6 & 7 & 8 & 9 &10 & 11 \\
\hline\hline
$(r_+/L)_{\textrm{crit}}$  &  0.8745 & 1.2665 & 1.5779  & 1.8454 & 2.0837 & 2.3006 & 2.5007 \\
\hline 
$(r_+/L)_{\textrm{GL}}$ & 0.8762 & 1.2689 & 1.5808 & 1.8486 & 2.0872 & 2.3041 & 2.5044
\end{tabular}
\end{center}
\caption{Critical value of $r_+/L$ obtained with our method (second row) compared to critical value that signals the onset of the GL instability (third row) as a function of the total number of spacetime dimensions $d$. The agreement improves  as $d$ increases.  }
\label{table:BS}
\end{table}

These results are rather surprising. Violation of the local Penrose inequality is a sufficient condition for instability, but not a necessary one. For any given choice of initial data, the value of $r_+/L$ at which ${\cal A}$ changes sign should be smaller than the value at which the GL instability appears, as we find. Only by considering a sufficiently general class of initial data would one expect our approach to be able to identify the critical value of $r_+/L$ exactly. Nevertheless, the very simple initial data we constructed using the conformal rescaling method gives a value of $r_+/L$ which is within $0.2 \%$ of the critical value. This suggests that our initial data is (at first order) quite close to initial data for the GL unstable mode with the same $x$-dependence as our perturbation.\footnote{More precisely, one can superpose the GL unstable mode, proportional to $e^{\Omega t}$, with its image under the $t \rightarrow -t$ isometry to obtain a time-symmetric unstable mode, which is more natural to compare with our initial data.} However, the GL unstable mode vanishes at the bifurcation surface whereas our initial data is non-vanishing there. Nevertheless, both perturbations are localized near the horizon so perhaps this is the reason for the surprising accuracy of our result.

Note that we needed to work to second order in perturbation theory to demonstrate instability even though the GL instability is present at first order. This will be true also in the other examples we study. The reason is that our argument involves the change in mass and horizon area sourced by the perturbation. In general, a first order perturbation can lead to a first order change in the mass, angular momentum, horizon area etc. but these changes are governed by the first law  \cite{Sudarsky:1992ty}. This implies that the local Penrose inequality is saturated at first order. Given a first order metric perturbation $\delta g_{ab}$, let $\delta g_{ab}^{ND}$ be a "non-dynamical", i.e., time-independent, perturbation that is obtained by a first order variation of the parameters of the unperturbed black hole solution, and has the same mass and angular momenta as $\delta g_{ab}$ at first order.  Write $\delta g_{ab}=\delta g_{ab}^{ND} + \delta g_{ab}^D$ where $\delta g_{ab}^D$ is "dynamical", i.e., time-dependent. If an instability is present then it is associated to $\delta g_{ab}^D$, not $\delta g_{ab}^{ND}$. But, by construction, $\delta g_{ab}^D$ makes only a second order contribution to the mass and angular momentum. Hence we have to work to second order in order to see the instability using a local Penrose inequality, even though this instability is present in the linearized theory.

\section{Rotating black holes}

\label{rotating}

\subsection{Penrose inequality}

Myers-Perry black holes are uniquely parameterized by their mass $M$ and angular momenta $J_i$ where $i=1,\ldots , N=[(d-1)/2]$. Black rings can be parameterized uniquely by $M$ and a pair of dimensionless parameters 
$(\nu,\alpha)$, which we shall define below. These quantities determine the angular momenta $J_1$, $J_2$. However, $(M,J_1,J_2)$ do not uniquely specify the ring. If we define
\be
\label{Jac}
 \Delta = \det \frac{\partial(M,J_1,J_2)}{\partial (M,\nu,\alpha)} =  \det \frac{\partial(J_1,J_2)}{\partial (\nu,\alpha)}
\ee
then we can divide rings into "thin" rings with $\Delta>0$ and "fat" rings with $\Delta < 0$, each of which is uniquely parameterized by $M$ and $J_i$. In discussing stability we can regard these as distinct families of solutions, e.g., a stable thin ring must remain on the thin ring branch if perturbed. 
 Let $A_{\rm BH}(M,J_i)$ denote the area of the event horizon for the particular family of black holes under consideration.

Consider a black hole belonging to one of these families and initial data corresponding to a small perturbation of the black hole. If the black hole is stable then, under time evolution, the perturbation should disperse, and the spacetime will settle down to a black hole belonging to the same family, with a small change in parameters. Let $M_F$ and $J_{iF}$ denote the final mass and angular momenta, and $A_I$ denote the area of the intersection of the event horizon with the initial data surface. As before, we define $A_{\rm min}$ to be the greatest lower bound on the area of any surface that encloses the apparent horizon on the initial data surface. Then
\be
\label{rotatingpenrose1}
 A_{\rm min} \le A_I \le A_{\rm BH}(M_F,J_{iF}) \le A_{\rm BH}(E,J_{iF})
\ee
where $E$ is the ADM energy of the initial data. The first inequality applies because the event horizon lies outside the apparent horizon, the second inequality is the second law and the final inequality uses $M_F \le E$ and the fact that $A_{\rm BH}$ is an increasing function of mass at fixed angular momenta (which follows from the first law). 

The inequality (\ref{rotatingpenrose1}) is not very useful because we know nothing about $J_{iF}$. However, Ref. \cite{gibbons} observed that one can circumvent this problem by imposing symmetries on the initial data which ensure that angular momentum is conserved. 

In our case, the families of solutions under consideration admit $N$ commuting rotational symmetries, for which the angular momenta $J_i$ are given by the associated Komar integrals. If we assume that our initial data preserves these rotational symmetries that these angular momenta will be conserved. Hence, for initial data corresponding to a small perturbation of the black hole, preserving the $N$ rotational symmetries, stability implies the local Penrose inequality
\be
\label{rotatingpenrose2}
 A_{\rm min} \le A_{\rm BH}(E,J_i)
\ee
where $J_i$ are the angular momenta of the initial data. 

This is saturated by a constant $t$ slice through the unperturbed black hole. Hence a stable black hole is a local maximum of horizon area at fixed mass and angular momentum, and a local minimum of mass at fixed horizon area and angular momentum, in the space of rotationally symmetric, asymptotically flat, initial data.

Finally, we have the problem that $A_{\rm min}$ is difficult to calculate. This is a problem that can be overcome by imposing an additional discrete symmetry on the initial data that ensures that the apparent horizon is a minimal surface. In 4d, an appropriate symmetry is "$t-\phi$ symmetry"  \cite{gibbons,hawking}. This means: the initial data $(\Sigma,h_{ab},K_{ab})$ is axisymmetric and  one can introduce coordinates on the initial surface such that (i) the Killing field associated to axisymmetry is $\Phi^a=(\partial/\partial \phi)^a$, and (ii) $\phi \rightarrow -\phi$ is a diffeomorphism which preserves $h_{ab}$ but reverses the sign of $K_{ab}$. A surface of constant $t$ in the Kerr spacetime has this symmetry. The existence of this symmetry implies that 
\be
 K_{ab} = 2 J_{(a} \Phi_{b)}
\ee
where $J_a$ is axisymmetric with $J_a \Phi^a = 0$ (and hence $K=0$, i.e., the slice is maximal). 

The analogous symmetry for $d>4$ dimensions, which we shall call "$t-\phi_i$ symmetry", assumes $N$ rotational symmetries, generated by $\Phi^{ia}=(\partial/\partial \phi_i)^a$, and that the diffeomorphism $\phi_i \rightarrow -\phi_i$ (simultaneously for all $i$) preserve $h_{ab}$ but reverses the sign of $K_{ab}$. Surfaces of constant $t$ in all of our black hole families possess this symmetry. This symmetry implies that
\be
\label{tphisymK}
 K_{ab} = 2  J^i_{(a} \Phi^i_{b)}
\ee
where $J^i_a$ is invariant under the rotational symmetries, and orthogonal to $\Phi^j_a$ for all $j$. 

Let $S$ denote the apparent horizon in initial data with this symmetry and let $n^a$ denote the outward unit normal to $S$ in $\Sigma$. The condition that $S$ is marginally outer trapped is
\be
 \hat{K} + (h^{ab} - n^a n^b) K_{ab} = 0,
\ee
where $\hat{K}$ is the trace of the extrinsic curvature of $S$ regarded as a surface in $\Sigma$. Now $S$ must be invariant under the rotational symmetries hence $n^a \Phi^i_a=0$. Using (\ref{tphisymK}) we then deduce that the apparent horizon satisfies $\hat{K}=0$, i.e., it is a minimal surface. 

Next we must show that the apparent horizon is a {\it stable} minimal surface, i.e., a local minimum of the area rather than just an extremum. For $d>4$ dimensions this has been established only for time-symmetric initial data \cite{Cai:2001su}. However, for $t-\phi_i$ symmetric initial data, we prove in  Appendix \ref{sec:apparent} that $S$ is a local minimum in the set of all homologous surfaces which lie outside $S$ and are tangent to the Killing fields $K^i_a$. Since the event horizon is such a surface, we can deduce that $A_{\rm app}$ (the apparent horizon area) is a lower bound for $A_I$ and hence
\be
\label{rotPenrose}
 A_{\rm app} \le A_{\rm BH}(E,J_i).
\ee
To summarize: if the black hole is stable then this inequality must be satisfied by asymptotically flat initial data describing a small perturbation of the black hole which preserves $t-\phi_i$ symmetry. $E$ is the ADM energy, and $J_i$ are the angular momenta, of the initial data.  The function $A_{\rm BH}$ is defined by the family of stationary black holes that one is considering.

\subsection{Initial data}

We will construct initial data describing a $t-\phi_i$ symmetric perturbation of the black hole by the Lichnerowicz method of conformal rescaling. We take the $t-\phi_i$ symmetric initial data $(\Sigma,\bar{h}_{ab},\bar{K}_{ab})$ on a constant $t$ slice of our black hole solution and rescale
\be
 h_{ab} = \Psi^{4/(d-3)} \bar{h}_{ab}, \qquad K_{ab} =\Psi^{-2} \bar{K}_{ab}
\ee 
where $\Psi$ is independent of the angles $\phi_i$. The momentum constraint is automatically satisfied. The Hamiltonian constraint reduces to
\be
 \bar{\nabla}^2 \Psi - \frac{(d-3)}{4(d-2)} \bar{R} \left( \Psi - \Psi^{-3-4/(d-3)} \right) = 0,
\ee
where $\bar{\nabla}$ is the connection, and $\bar{R}$ the Ricci scalar, associated to $\bar{h}_{ab}$. 

Let $U$ denote a surface homologous to the apparent horizon of the unperturbed black hole initial data, and lying a finite distance behind the horizon. Let $\Sigma'$ denote the region of $\Sigma$ exterior to $U$, so $\partial \Sigma' = U\cup S_\infty$ where $S_\infty$ denotes a sphere at spatial infinity. We will solve the above equation on $\Sigma'$. For small enough initial data, the apparent horizon will be close to that of the unperturbed black hole and therefore it will lie outside $U$ and hence on $\Sigma'$. 

Now we consider boundary conditions. We will demand that $\Psi = 1 + {\cal O}(r^{-3+d})$ at $S_\infty$, which ensures asymptotic flatness. On $U$ we will impose Dirichlet boundary conditions, i.e., we will specify $\Psi|_U$. We will choose this function according to the kind of instability that we expect. The Hamiltonian constraint for the unperturbed black hole implies that $\bar{R} \ge 0$, which guarantees that there will be at most one solution of the above equation satisfying these boundary conditions. 

It would be straightforward to solve this nonlinear problem numerically. However, since we are interested in arbitrarily small initial data, we will proceed perturbatively. We seek a 1-parameter family of  solutions $\Psi(\lambda)$ with $\Psi(0)=1$ (we suppress the dependence of $\Psi$ on the coordinates). We then Taylor expand $\Psi(\lambda) = 1 + \lambda \dot{\Psi}(0) + (1/2) \lambda^2 \ddot{\Psi}(0) + \ldots$ where a dot denotes a derivative with respect to $\lambda$. Substituting into the above equation gives
\be
\label{Psi1}
 \bar{\nabla}^2 \dot{\Psi} -  \bar{R} \dot{\Psi} = 0,
\ee
and
\be
\label{Psi2}
 \bar{\nabla}^2 \ddot{\Psi} -  \bar{R} \ddot{\Psi} = - \left( \frac{3d-5}{d-3} \right)\,\bar R\, \dot{\Psi}^2
\ee
with evaluation at $\lambda=0$ understood. We solve (\ref{Psi1}) subject to the boundary conditions $\dot{\Psi}={\cal O}(r^{-d+3})$ at $S_\infty$ and $\dot{\Psi} = \dot{\Psi}|_U$ on $U$ for suitably chosen $\dot{\Psi}|_U$. We then solve (\ref{Psi2}) with the boundary conditions $\ddot{\Psi}={\cal O}(r^{-d+3})$ at $S_\infty$ and $\ddot{\Psi} = 0$ on $U$.\footnote{This could be generalized by taking $\ddot{\Psi}$ to be some specified function on $U$. The resulting solution would differ from our solution by a function $\ddot{\Psi}_2$ satisfying the same homogeneous equation as $\dot{\Psi}$. However the contribution from $\ddot{\Psi}_2$ to the Penrose inequality expanded to ${\cal O}(\lambda^2)$ is equivalent to replacing $\dot{\Psi}$ with $\dot{\Psi} + (1/2) \lambda \ddot{\Psi}_2$ in the ${\cal O}(\lambda)$ terms. Since the latter quantity satisfies the same equation as $\dot{\Psi}$, the first law will ensure that these terms drop out of the Penrose inequality (see main text). Hence including $\ddot{\Psi}_2$ has no effect and there is no loss of generality in our choice of boundary condition for $\ddot{\Psi}$.} Our problem therefore is specified by the choice of the function $\dot{\Psi}|_U$. 

The boundary conditions at $S_\infty$ imply that our initial data has the same angular momentum $J$ as the unperturbed black hole solution, i.e., $J(\lambda)=J$. The ADM momentum vanishes, so $E=M$, the ADM mass. 
The mass $M(\lambda)$ is determined by $M(0)=M$ (the mass of the unperturbed black hole) and
\be
\label{dotM}
 \dot{M}(0) = \frac{(d-2) \Omega_{d-2}}{4 \pi} \left( r^{d-3} \dot{\Psi}\right)|_{r=\infty}, \qquad \ddot{M}(0) = \frac{(d-2) \Omega_{d-2}}{4 \pi} \left( r^{d-3} \ddot{\Psi}\right)|_{r=\infty}
\ee
The first law (proved for $t-\phi$ symmetric data in \cite{Hawking:1973aq} and general initial data in \cite{Sudarsky:1992ty}) guarantees that the first order change in the apparent horizon area satisfies
\be
\label{1stlaw}
 \frac{1}{4} T \dot{A}_{\rm app}(0) = \dot{M}(0)
\ee
where $T$ is the temperature of the unperturbed black hole. The calculation of the second order change is described in Appendix \ref{sec:horizonarea}. Next we expand the RHS of (\ref{rotPenrose}) to second order in $\lambda$ using
\bea
 \left( \frac{d^2}{d\lambda^2} A_{\rm BH}(M(\lambda),J) \right)_{\lambda=0} &=& \frac{\partial A_{\rm BH}}{\partial M}(M(0),J) \ddot{M}(0) + \frac{\partial^2 A_{\rm BH}}{\partial M^2}(M(0),J) \dot{M}(0)^2 \nonumber \\
&=& \frac{4}{T} \ddot{M}(0) - \frac{4}{T^2 c_J} \dot{M}(0)^2 
 \eea
where
\be
 c_J = \left(\frac{\partial M}{\partial T}\right)_J
\ee
is the heat capacity at constant angular momentum of the unperturbed black hole. The first order terms in the local Penrose inequality (\ref{rotPenrose}) cancel using the first law. At second order it becomes (for small enough $\lambda$)
\be
 Q \ge 0
\ee
where
\be
\label{Qdef}
 Q \equiv \ddot{M}(0) - \frac{1}{4} T \ddot{A}_{\rm app}(0) - \frac{1}{T c_J} \dot{M}(0)^2
\ee
The Penrose inequality is violated for arbitrarily small $\lambda$ if $Q < 0$. Our strategy for demonstrating instabilities will be to seek a function $\dot{\Psi}|_U$ such that $Q<0$. 

Note the presence of $c_J$ in the denominator of the final term of (\ref{Qdef}). What happens if $c_J$ vanishes for some particular black hole solution? This does not happen for the Myers-Perry solution. However, some black rings do have vanishing $c_J$. In fact, "thin" rings have $c_J<0$ and "fat" rings have $c_J>0$, with $c_J$ passing through zero as one moves from the thin to the fat branch.\footnote{\label{DeltacJ}
A plot of temperature against mass at fixed angular momenta has a vertical asymptote at a point where $c_J=0$. Near the asymptote there will be two solutions with the same mass and angular momenta, these are the thin and fat rings.}  For thin rings, the final term in (\ref{Qdef}) tends to make $Q$ more positive. For fat rings it makes $Q$ more negative. In particular, for a fat ring with very small $c_J$ it seems that $Q$ will be negative as long as the initial data has $\dot{M}(0) \ne 0$. This strongly suggests that such a ring will be unstable. 

This is very similar to the argument used in Ref. \cite{Arcioni:2004ww} to deduce that fat black rings are unstable near the "turning point" where $c_J=0$. However, in place of the unspecified manifold ${\cal M}$ of Ref. \cite{Arcioni:2004ww}, we are working in the more well-defined setting of a 1-parameter family of initial data for Einstein's equation. This makes it clear that the instability (if one exists) is classical. Furthermore, the maximization of entropy argument of Ref. \cite{Arcioni:2004ww} is given a more precise formulation by the local Penrose inequality for rotationally symmetric initial data.\footnote{Ref. \cite{Arcioni:2004ww} was uncertain about whether the "turning point" method requires rotational symmetry. We believe it to be essential for without it one cannot obtain a local Penrose inequality, which is the reason for believing that a stable black hole must be a local maximum of horizon area for fixed mass and angular momentum.} 

Finally, note that this argument is not a {\it proof} that fat black rings with small $c_J$ are unstable because we have not excluded the possibility that other terms in (\ref{Qdef}) also become large. To demonstrate instability we will have to construct initial data for which $Q<0$.

\section{Myers-Perry black holes}

\subsection{Background solution}

We will consider a singly spinning Myers-Perry black hole with metric
\be
 ds^2=-\frac{\rho^2\Delta}{\Sigma^2}dt^2+\frac{\Sigma^2\sin^2\theta}{\rho^2}
\left(d\phi-\Omega dt\right)^2
+\frac{\rho^2}{\Delta}dr^2+\rho^2d\theta^2+r^2\cos^2\theta d\Omega_{(d-4)}^2\ ,
\ee
where
\begin{equation}
\begin{split}
&\Delta=r^2+a^2-\frac{r_M^{d-3}}{r^{d-5}}\ ,\quad
\rho^2=r^2+a^2\cos^2\theta\ ,\\
&\Sigma^2=(r^2+a^2)^2-a^2\Delta\sin^2\theta\ ,\quad
\Omega=\frac{r_M^{d-3} a}{\Sigma^2 r^{d-5}}
\end{split}
\end{equation}
The solution depends on the two parameters $r_M$ and $a$. 
The horizon of this black holes is located at $r=r_+$: the largest root of $\Delta(r)=0$. For $d=4$ and $d=5$, the rotation parameter $a$ have the bound $a^2\leq r_M^2/4$
and $a^2< r_M^2$, respectively, with a strict inequality in the $d=5$ case since we want to consider only regular spacetimes. For $d\geq 6$, there is no bound for the
rotation parameter. The symmetry of this spacetime is $R_t \times U(1) \times SO(d-3)$,
where $R_t$ is the time translation symmetry, $U(1)$ is the rotational
symmetry generated by $\partial_\phi$ and $SO(d-3)$ is the symmetry of
$d\Omega_{(d-4)}^2$ part of the metric.
The thermodynamical parameters are
\begin{equation}
\begin{split}
&M=\frac{(d-2)\Omega_{d-2}r_M^{d-3}}{16\pi }\ ,\quad
 J=\frac{2}{d-2}Ma\ ,\quad
A_H=\Omega_{d-2}r_+^{d-4}(r_+^2+a^2)
\\
&T=\frac{(d-3)r_+^2 + (d-5)a^2}{4\pi r_+(r_+^2+a^2)}\ ,\quad
\Omega_H = \frac{a}{r_+^2+a^2}\ ,
\end{split}
\end{equation}
where $M$, $J$, $A_H$, $T$ and $\Omega_H$ are ADM mass. ADM angular
momentum, area of the event horizon, Hawking temperature and angular
velocity of horizon, respectively. $\Omega_{d-2}$ is the area of
a unit $(d-2)$-sphere.
The heat capacity at constant angular momentum is
\begin{equation}
 c_J=-\frac{1}{4}\frac{(d-2)\Omega_{d-2}r_+^{d-4}(r_+^2+a^2)^2\{(d-3)r_+^2+(d-5)a^2\}}{(d-3)r_+^4-6r_+^2
  a^2 +3(d-5)a^4}\ .
\end{equation}
This quantity is negative for small $a/r_+$ and  it never vanishes. However, for $d=5$ it diverges (and changes sign) at $a/r_+ = 1/\sqrt{3}$. For $d=6$ it diverges at $a/r_+=1$ but it does not change sign. For $d\ge 7$ it is finite and negative for all $a/r_+$. Note that a divergence (and sign change) also occurs for the Kerr black hole. 

Consider a surface of constant $t$ in this geometry. To extend it through the Einstein-Rosen bridge, we define a new "radial" coordinate $z$ by
\be
 z^2=\frac{r-r_+}{r_+}\ .
\ee
The resulting surface $\Sigma$ has geometry
\be
 ds^2 =\frac{\Sigma^2\sin^2\theta}{\rho^2}
d\phi^2
+\frac{4r_+^2z^2 \rho^2}{\Delta}dz^2+\rho^2d\theta^2+r^2\cos^2\theta d\Omega_{(d-4)}^2\ .
\ee
Note that near the horizon  we have $\Delta/z^2\simeq \Delta'(r_+)r_+^2$ and 
therefore, above metric is regular at $z=0$ (the horizon). In these new coordinates the region of $z<0$
corresponds to the opposite side of the Einstein-Rosen bridge and the two asymptotically flat regions are related by the isometry $z\leftrightarrow -z$. 

The extrinsic curvature of this surface has the form (\ref{tphisymK}) with just a single non-vanishing $J^i_a$, corresponding to the rotational Killing vector $\Phi^a = (\partial/\partial \phi)^a$:
\be
 K_{ab} = 2J_{(a} \Phi_{b)}, \qquad J_a =
 -\frac{\Sigma}{2\rho\sqrt{\Delta}}\partial_a\Omega
\ee
One can check that $J_a$ is smooth at $z=0$.

\subsection{Numerical strategy}

Now, we explain the numerical strategy to find the ultra-spinning
instability of 
Myers-Perry black holes.
First of all, we have to solve equations for the conformal
factor~Eqs.(\ref{Psi1}) and (\ref{Psi2}).
It is convenient to define new variables as
\begin{equation}
 \dot{\psi}\equiv \left(\frac{r}{r_+} \right)^{d-3} \dot{\Psi}\ ,\qquad
 \ddot{\psi}\equiv \left( \frac{r}{r_+} \right)^{d-3} \ddot{\Psi}\ .
\label{fac}
\end{equation}
From the asymptotic flatness, variables $\dot{\Psi}$ and $\ddot{\Psi}$
must decay as $\mathcal{O}(r^{-d+3})$ at infinity and, thus, we impose Neumann
boundary conditions for $\dot{\psi}$ and $\ddot{\psi}$ at infinity.
Then, from asymptotic values of  $\dot{\psi}$ and
$\ddot{\psi}$, we can easily determine the deviation of mass
using Eq.(\ref{dotM}). 

We also have to impose boundary conditions at an inner boundary.
Without loss of generality we will choose the surface $U$ defined above to be located at $z=-0.5$, which is
well behind the horizon ($z=0$) of the unperturbed spacetime.
Any smooth scalar in this geometry must be an even function of $\theta$ near $\theta=0$ and an even function of $\pi/2-\theta$ near $\theta=\pi/2$. Such a function can be extended to an even function of $\theta$ with period $\pi$ and can therefore be expanded in a Fourier cosine series with terms $\cos 2n\theta$. Therefore we choose our boundary condition on $U$ to be
\be
 \dot{\psi}|_U = 1+\sum_{n=1}^{N}c_n\cos2n\theta\ ,
\label{Fex}
\ee 
where $c_n$ $(n=1,2,\cdots,N)$ are constants and the integer $N$
represent a truncation of the Fourier expansion. We have fixed the normalization of
$\dot{\psi}$, by choosing the first term in the expansion~(\ref{Fex}) to be unity.

As explained above, we will set $\ddot{\psi}|_U=0$. We then solve equations \eqref{Psi1}-\eqref{Psi2}) numerically. The smoothness conditions just mentioned require that we impose Neumann boundary conditions at $\theta=0,\pi/2$:
\begin{equation}
 \partial_\theta \dot{\psi}|_{\theta=0}
= \partial_\theta \ddot{\psi}|_{\theta=0}
=\partial_\theta \dot{\psi}|_{\theta=\pi/2}
= \partial_\theta \ddot{\psi}|_{\theta=\pi/2}=0\ .
\end{equation}
Our task is to choose the constants $N$ and $c_n$ appropriately to find an instability. We shall take $N=4$ and choose the constants $c_n$ to minimize (numerically) the quantity $Q$ defined in Eq.(\ref{Qdef}).

In order to suppress the truncation error at infinity in our numerical calculation we use a compact
radial coordinate $w$ defined as
\begin{equation}
 z=\tan\left(\frac{\pi}{2}w\right)\ .
\end{equation}
The coordinate range of $w$ is $\tan^{-1}(-0.5)\leq w < 1$.
Discretizing Eqs.(\ref{Psi1}) and (\ref{Psi2}) in the $w$-coordinate,
we obtain systems of linear equations.
We solved these equations using two different methods:
conjugate gradient and successive over-relaxation \cite{numerical}.
We found that these methods gave same result within numerical error.

\begin{figure}[htbp]
  \centering
  \subfigure
  {\includegraphics[height=5.0cm,clip]{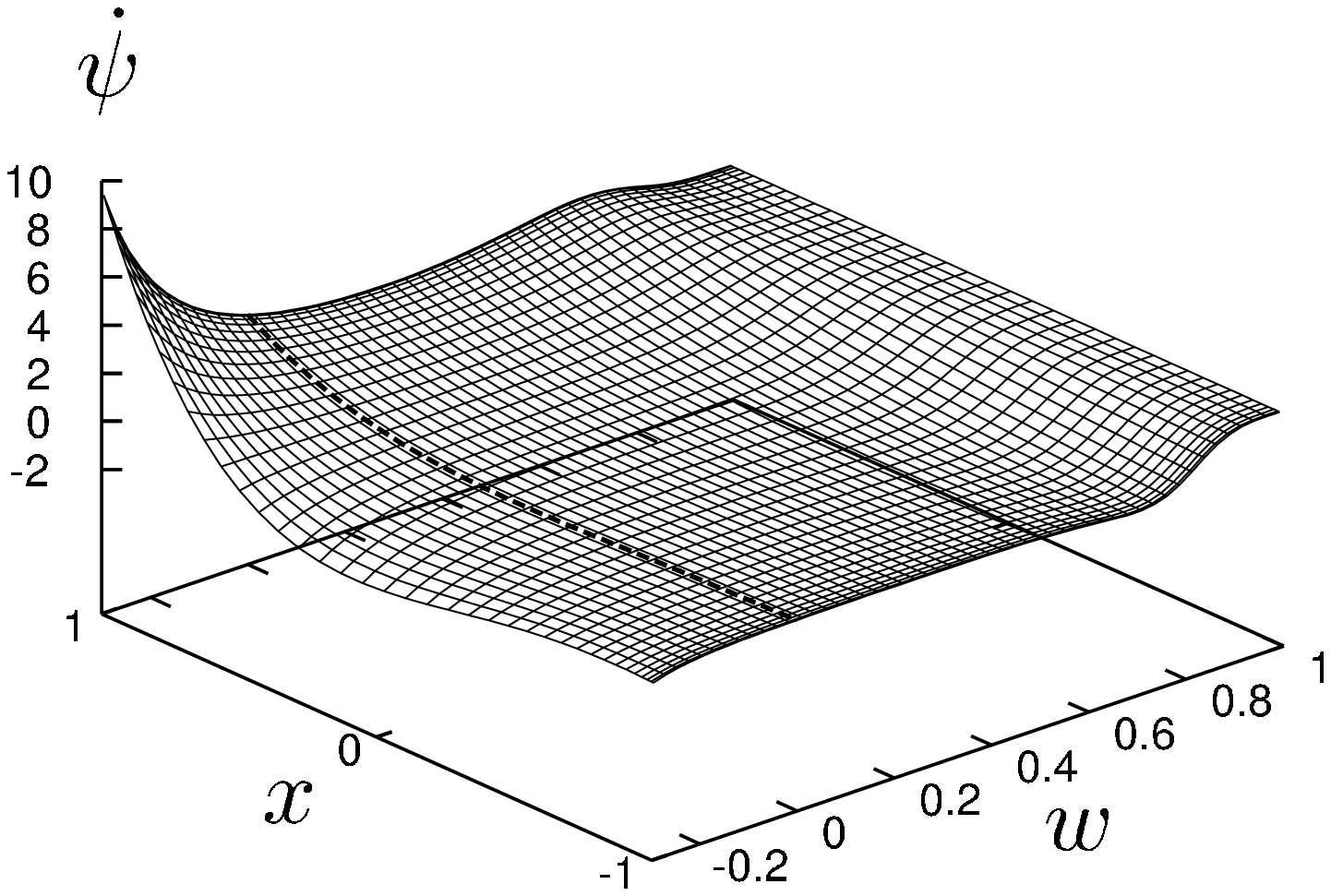}
    \label{fig:dpsi}
  }
  \subfigure
  {\includegraphics[height=5.0cm,clip]{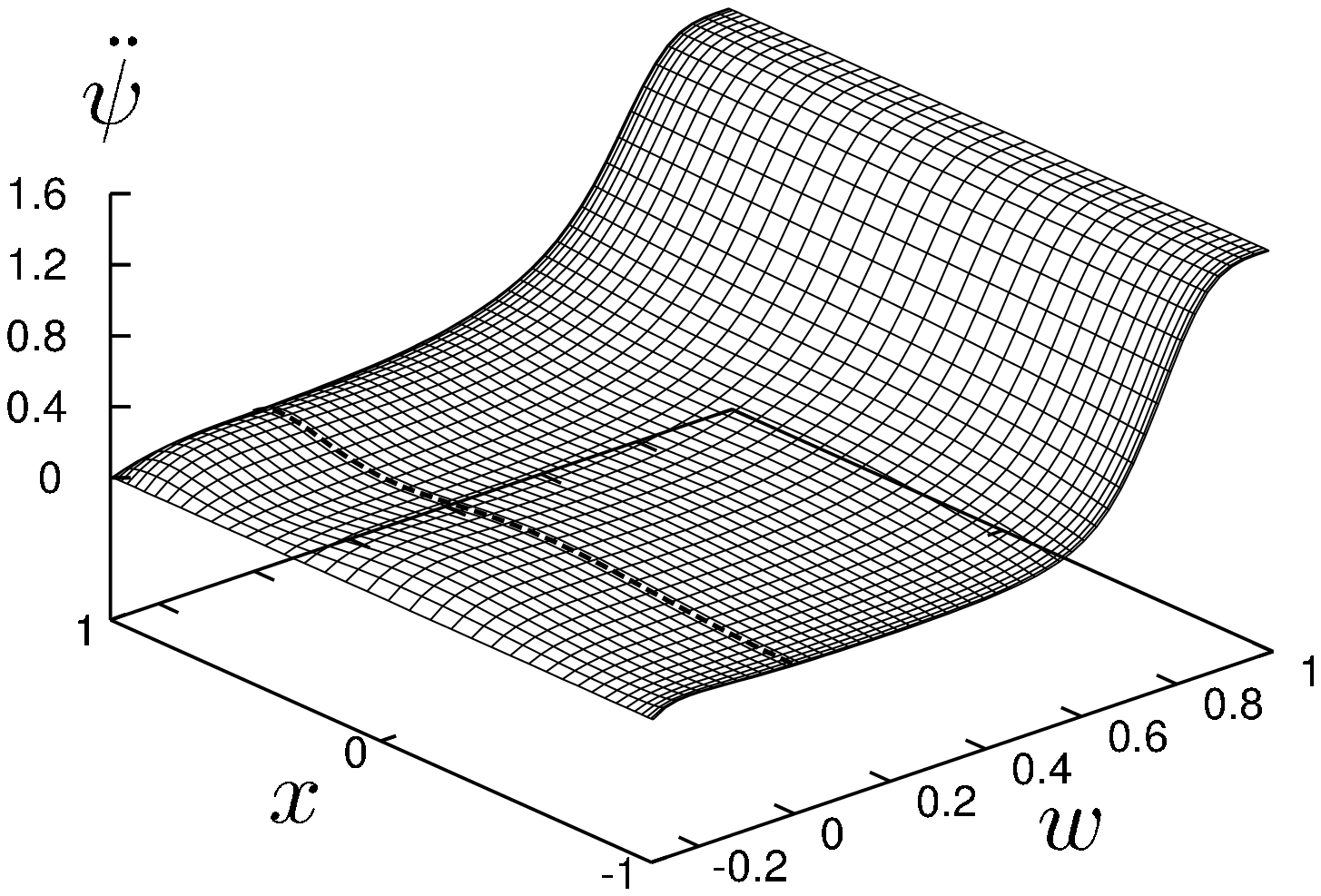} 
    \label{fig:ddpsi}
  }
  \caption{\label{fig:psis}
The solutions $\dot{\psi}$ (left) and $\ddot{\psi}$ (right) for
 $a/r_M=2$ and $(c_1,c_2,c_3,c_4)=(3.51,3.12,1.50,0.239)$. The dotted line shows the position of the horizon of the unperturbed solution ($w=0$).
}
\end{figure}

In Figure. \ref{fig:psis}, as an example, we show the solution for  $d=6$, $a/r_M=2$,\footnote{
The background solution can be parameterized with $r_M$ and $a/r_M$. The former is dimensionful and so it just defines a scale. Therefore we shall just state the value of the dimensionless parameter $a/r_M$ when giving results.} and $(c_1,c_2,c_3,c_4)=(3.51,3.12,1.50,0.239)$ (These values
of $c_n$'s were chosen to minimize $Q$.).
Since we have defined $\dot{\psi}$ and $\ddot{\psi}$ as in
Eq.(\ref{fac}), these functions approach constant values at infinity
$w\to 1$. From the asymptotic values of $\dot{\psi}$ and $\ddot{\psi}$,
we can determine the deviation of mass $\dot{M}(0)$ and $\ddot{M}(0)$
using Eq. (\ref{dotM}). 
From Eq. (\ref{Adotzero}) of Appendix \ref{sec:horizonarea}, we can  
obtain the first order deviation of the area of the apparent horizon.\footnote{
Note that $z$ plays the role of the coordinate $r$ used in Appendix \ref{sec:horizonarea}.}
The explicit expression is
\begin{equation}
\dot{A}_\textrm{app}(0)=
\int_{-1}^1 dx\,I_0(x)\dot{\Psi}(z=0,x) 
\ .
\end{equation}
where we have defined
\be
 x = \cos 2\theta
\ee
and
\begin{equation}
I_0=\frac{(d-2)\pi\Omega_{d-4}r_+^{d-4}(r_+^2+a^2)}{d-3}
\left(\frac{1+x}{2}\right)^{\frac{d-5}{2}}\ .
\label{I0}
\end{equation}
The first order changes in the mass and horizon area must satisfy the first law (\ref{1stlaw}) (recall that our perturbation does not change the angular momentum) and therefore we can use it as a global measure of the numerical error of our calculations. In the following we only present data for which  the numerical error is less than $1\%$ (see Appendix \ref{sec:Convergence} for the details and the convergence tests).

\begin{figure}[htbp]
  \begin{center}
   \includegraphics[height=6.0cm,clip]{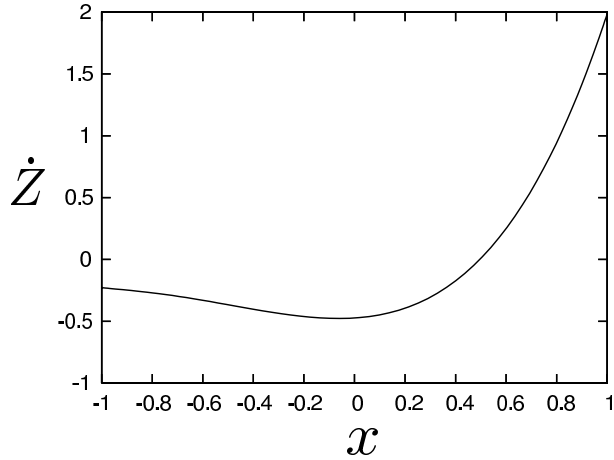}
   \caption{
\label{fig:zah}
The first order deviation of the apparent horizon for the solution of Fig. \ref{fig:psis}.
}
  \end{center}
\end{figure}

To evaluate the second order deviation of the area of the apparent
horizon, we need to find the apparent horizon in the perturbed initial
data.  Let the first order position of the apparent horizon be $z = \lambda \dot{Z}(x)$.
From Eq.(\ref{deltaEL}) of Appendix \ref{sec:horizonarea}, we have the equation 
\begin{equation}
 \frac{d^2\dot{Z}}{dx^2}+A(x) \frac{d\dot{Z}}{dx}+ B(x) \dot{Z} + C(x) \partial_z \dot{\Psi}(z=0,x) = 0
\label{appeq}
\end{equation}
where functions $A$, $B$ and $C$ are defined as
\begin{equation}
\begin{split}
&A=\frac{(d-5)-(d-1)x}{2(1-x^2)}\ ,\\
&B=-\frac{(d-3)r_+^2+(d-5)a^2}{32r_+^2(r_+^2+a^2)^2(1-x^2)}\big[
\{(d-3)r_+^2+(d-5)a^2\}a^2 x \\
&\hspace{3cm}+
\{4(d-2)r_+^4+7(d-3)r_+^2a^2+(3d-11)a^4\}
\big]\ ,\\
&C=-\frac{2(d-2)}{d-3}\frac{(d-3)r_+^2+(d-5)a^2}{16r_+^2(1-x^2)}\ .
\end{split}
\end{equation}
Regularity at $x=\pm 1$ gives the boundary condition that $\dot{Z}$ should be finite at $x = \pm 1$. 
Using the solution for $\dot{\psi}$ shown in Fig. \ref{fig:psis}, we find the location of the apparent horizon shown in Fig. \ref{fig:zah}.
From the solution of $\dot{Z}(x)$, we can calculate the second order deviation of the
area of the apparent horizon. From Eq.(\ref{ddotA}) of Appendix \ref{sec:horizonarea}, we obtain the
explicit expression for $\ddot{A}_\textrm{app}$ as 
\begin{equation}
\label{darea}
\ddot{A}_\textrm{app}=\int^1_{-1}dx\,\bigg[
I_0(x)\left(
2\partial_z\dot{\Psi}\dot{Z}
+\ddot{\Psi}
+\frac{d-1}{d-3}\dot{\Psi}^2
\right)_{z=0}
+I_1(x)\left( \frac{d \dot{Z}}{dx} \right)^2+I_2(x)\dot{Z}^2\bigg]\ ,
\end{equation}
where
\begin{equation}
\label{ddarea}
\begin{split}
&I_1(x)=4\pi\Omega_{d-4} 
 \frac{4r_+^{d-2}(r_+^2+a^2)(1-z)}{(d-3)r_+^2+(d-5)a^2}\left(\frac{1+x}{2}\right)^{\frac{d-3}{2}}\
 , \\
&I_2(x)=4\pi\Omega_{d-4}\frac{r_+^{d-4}}{16(r_+^2+a^2)}\left(\frac{1+x}{2}\right)^{\frac{d-5}{2}}
\big[
\{(d-3)r_+^2+(d-5)a^2\}a^2 x \\
&\hspace{5cm}+
\{4(d-2)r_+^4+7(d-3)r_+^2a^2+(3d-11)a^4\}
\big]\ ,
\end{split}
\end{equation}
and $I_0(x)$ was defined in Eq.(\ref{I0}).
Substituting the values for $\dot{M}$, $\ddot{M}$, $\dot{A}_\textrm{app}$ and $\ddot{A}_\textrm{app}$
into Eq.(\ref{Qdef}), we can evaluate $Q$.

\subsection{Results}

Using above procedure, we can evaluate $Q$ for each choice of $N$ and $\{c_n \}$.
To find initial data which satisfies $Q<0$, we seek to determine the  $\{c_n\}$ which minimize the dimensionless quantity
\begin{equation}
 \bar{Q}=\frac{Q}{D^2 M}\ , \qquad D^2=\int^{\pi/2}_0 d\theta \, \dot{\Psi}(z=0,\theta)^2
\end{equation}
The factor of $D^2$ is included because $Q$ will scale as the square of the amplitude of $\dot{\Psi}$. We have fixed the normalization of $\dot{\Psi}$ by equation (\ref{Fex}) but the $c_n$ might become large in the minimization process. Dividing by $D^2$ reduces the chance of the minimization algorithm running off to large $c_n$. 

\begin{figure}[htbp]
  \begin{center}
   \includegraphics[height=6.0cm,clip]{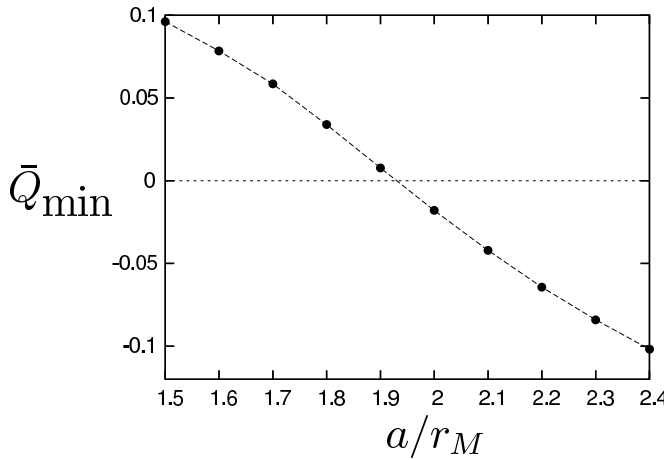}
   \caption{
\label{fig:Qmin} The minimum value of $\bar{Q}$ is plotted against $a/r_M$.
 $\bar{Q}_\textrm{min}$ is negative for $a/r_M>1.933$.
}
  \end{center}
\end{figure}

To carry out the minimization numerically we have used  the downhill simplex method of Nelder and Mead \cite{Nelder} (see also \cite{numerical}) setting 
$N=4$ in Eq. (\ref{Fex}).\footnote{ We have checked that the result does not change much for larger $N$.}
In Fig. \ref{fig:Qmin}  we plot $\bar{Q}_\textrm{min}$ against $a/r_M$ for $d=6$.
We can see that the $\bar{Q}_\textrm{min}$ is negative for $a/r_M>1.933$, which proves that singly spinning 6d MP black holes with $a/r_M>1.933$ are unstable. 

Ref.~\cite{Dias:2010maa} found that a time-independent mode indicating the onset of instability appears at  $a/r_M=1.572$ so the instability should be present for $a/r_M>1.572$. This emphasizes that our approach gives a sufficient condition for instability but not a necessary condition. If we considered a more general class of initial data, which would involve going beyond the conformal rescaling method used above, then we should be able to get closer to the bound of Ref.~\cite{Dias:2010maa}. However, it is striking that our very simple approach yields results that are so close to the exact onset of instability.

We have repeated the above calculation for other values of $d$. For $d=5$ we find that $Q$ is always positive. Therefore our results are consistent with the stability of $d=5$ single spinning Myers-Perry black holes against rotationally symmetric perturbations, in agreement with Refs. \cite{Dias:2009iu,Dias:2010maa}. (However, Ref. \cite{Shibata:2009ad} found that there is an instability which breaks rotational symmetry.) In Table \ref{onset}, we summarize the our results for $6 \le d \le 11$. We give the critical value $a/r_M$ beyond which we predict instability and the corresponding results of Dias et al~\cite{Dias:2010maa}. 

Our approach becomes less good at identifying the onset of instability as $d$ increases. So presumably 
our (first order) initial data looks less like initial data for an unstable mode for larger $d$.
\begin{table}[h]
\begin{equation}
\begin{array}{ c| c |c| c| c| c| c}
d & 6 & 7 & 8 & 9 & 10 & 11 \\ \hline\hline
a/r_M & 1.933 & 2.380 & 2.635 & 2.803 & 2.934 & 3.048\\
\hline
a/r_M(\textrm{Dias et al.}) & 1.572 & 1.714 & 1.770 & 1.792 & 1.795 & 1.798\\
\end{array}
\nonumber
\end{equation}
\caption{\label{onset} 
Second row: critical value of $a/r_M$ at which the local Penrose inequality is violated.
Third row: onset of the instability found in Ref.~\cite{Dias:2010maa}.
}
\end{table}

\section{Singly spinning black rings}

\subsection{Background solution}

The metric of a singly spinning black ring \cite{Emparan:2001wn} in the coordinates of Ref. \cite{Emparan:2004wy} is
\begin{multline}
\label{neutral}
ds^2=-\frac{F(y)}{F(x)}\left(dt-C R\frac{1+y}{F(y)}
d\phi_2\right)^2 \\
+\frac{R^2}{(x-y)^2}F(x)\left[
-\frac{G(y)}{F(y)}d\phi_2^2-\frac{dy^2}{G(y)}
+\frac{dx^2}{G(x)}+\frac{G(x)}{F(x)}d\phi_1^2\right]\,,
\end{multline}
where
\bea\label{fandg}
F(\xi)=1+\lambda\xi,\qquad G(\xi)=(1-\xi^2)(1+\nu\xi),\qquad
C=\sqrt{\lambda(\lambda-
\nu)\frac{1+\lambda}{1-\lambda}}\,.
\eea
The constant $R$ has dimensions of length and sets a scale for the solution. The dimensionless parameters $\lambda$ and $\nu$ must lie in the range 
\bea\label{lanurange}
0< \nu\leq\lambda<1\,.
\eea 
Absence of conical singularities fixes $\lambda$ and the periodicity of $\phi_1,\phi_2$:
\begin{equation}
 \lambda=\frac{2\nu}{1+\nu^2}\ ,\quad
 0\leq \phi_1,\phi_2 \leq 2\pi\frac{\sqrt{1-\lambda}}{1-\nu}\ .
\end{equation}
The ranges of the coordinates $x$ and $y$ are $-1\leq x\leq1$ and $-1/\nu\leq y<-1$.
The event horizon is located at $y=-1/\nu\equiv y_h$.
The mass, angular momentum, horizon area, temperature, and angular velocity are
\begin{equation}
\begin{split}
&M=\frac{3\pi R^2}{2}\frac{\nu}{(1-\nu)(1+\nu^2)}\ ,\quad
J=\frac{\pi \nu R^3}{\sqrt{2}}
\left(
\frac{1+\nu}{(1-\nu)(1+\nu^2)}
\right)^{3/2}\ ,\\
&A_H=\frac{8\sqrt{2}\pi^2 \nu^2 R^3}{(1-\nu)(1+\nu^2)^{3/2}}\ ,\quad
T=\frac{(1-\nu)\sqrt{1+\nu^2}}{4\sqrt{2}\pi R \nu}\ ,\quad
\Omega_H=\frac{1}{R}\sqrt{\frac{(1-\nu)(1+\nu^2)}{2(1+\nu)}}\ .
\end{split}
\end{equation}
The heat capacity at constant angular momentum is given by
\begin{equation}
 c_J=\frac{12\sqrt{2}\pi^2(\nu-1/2)\nu^2\sqrt{1+\nu^2}}{(1-\nu)(2+\nu^2)(1+\nu^2)^2}R^3\ .
\end{equation}
A "thin" ring has $0 < \nu < 1/2$ and $c_J<0$. A "fat" ring has $1/2 < \nu< 1$ and $c_J>0$. A ring with $\nu=1/2$ is called "minimally spinning" because it has the minimum $J$ for given $M$.

Consider the induced metric on a constant $t$ surface $\Sigma$. There is a coordinate singularity at the bifurcation surface $y=y_h$. This can be eliminated by defining 
\be
 z^2=y-y_h\ .
\ee
In the $z$ coordinate, the induced metric $\bar{h}_{ab}$ on $\Sigma$ can be written as
\begin{equation}
\label{iniBR}
ds^2= 
\frac{R^2}{(x-y)^2}F(x)\left[
-\frac{G(y)}{F(y)}d\phi_2^2-\frac{4z^2dz^2}{G(y)}
+\frac{dx^2}{G(x)}+\frac{G(x)}{F(x)}d\phi_1^2\right]-\frac{C^2 R^2(1+y)^2}{F(x)F(y)}d\phi_2^2\,
\,,
\end{equation}
where $y=y(z)$. Near $z=0$ we have $G(y)/z^2\simeq G'(y_h)$ and so this metric can be smoothly extended to a new asymptotically flat region with negative $z$. This is related to the original region by the isometry $z \rightarrow -z$. The coordinate range of $z$ is then $-z_\textrm{max}<z<z_\textrm{max}$ where
\be
z_\textrm{max}=\sqrt{-1-y_h}
\ee

\subsection{Numerical strategy for singly spinning black rings}

\begin{figure}[htbp]
  \begin{center}
   \includegraphics[height=6.0cm,clip]{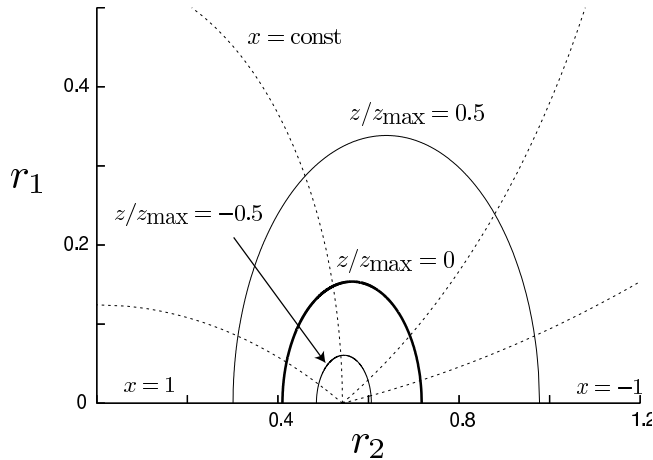}
   \caption{
\label{fig:r1r2} Adapted coordinates $(r_1,r_2)$ for $R=1$ and $\nu=0.6$.
The thick curve $z/z_\textrm{max}=0$ corresponds to the event horizon of
the background solution. The region inside the thick curve corresponds
   to the opposite side of the Einstein-Rosen bridge.
We impose a Dirichlet boundary condition at $z/z_\textrm{max}=-0.5$ which is located
   behind the horizon $z/z_\textrm{max}=0$.
}
  \end{center}
\end{figure}

In the  ring coordinates $x$ and $y$ (or $z$), asymptotic infinity
corresponds to a single point $x\to y\to -1$, which makes it difficult to impose boundary conditions
there. Therefore we introduce new coordinates
 $(r_1,r_2)$ defined by
\begin{equation}
 r_1=\tilde{R}\frac{\sqrt{1-x^2}}{x-Y(z)}\ ,\quad
 r_2=\tilde{R}\frac{\sqrt{Y(z)^2-1}}{x-Y(z)}\ ,
\label{r1r2def}
\end{equation}
where
\begin{equation}
 Y(z)=-\frac{(1-4z_\textrm{max}^2)z+z_\textrm{max}(1+4z_\textrm{max}^2)}{z+z_\textrm{max}}\ , \qquad \tilde{R}=R \sqrt{\frac{1-\lambda}{1-\nu}}
\end{equation}
The function $Y(z)$ is chosen
to satisfy $Y(z)\simeq y$ for $z\to z_\textrm{max}$ and $Y(z)\to -\infty$ for $z\to -z_\textrm{max}$.
In these coordinates, spatial infinity corresponds to 
$r_1\to \infty$ or $r_2\to \infty$ and the asymptotic form of the metric
is
\begin{equation}
 ds^2\simeq dr_1^2 + r_1^2 d\tilde{\phi_1}^2 + dr_2^2 + r_2^2 d\tilde{\phi_2}^2\,,
\label{BRasym}
\end{equation}
where $(\tilde{\phi_1},\tilde{\phi_2})=(1-\nu)/\sqrt{1-\lambda}(\phi_1,\phi_2)$
so that the periodicity of $\tilde{\phi_1}$ and $\tilde{\phi_2}$ is $2\pi$.
In Figure \ref{fig:r1r2}, we depict the relation between $(r_1,r_2)$ and
$(x,z)$ coordinates. Note that these new coordinates treat the two asymptotic regions asymmetrically, with asymptotic infinity in the negative $z$ region corresponding to the point $r_1=\tilde{R}$, $r_2=0$. This is not a problem because we will not be solving any equations near this point.

Now  we consider the equations for the conformal factor~(\ref{Psi1}) and
(\ref{Psi2}). Requiring asymptotic flatness implies that $\dot{\Psi}$ and $\ddot{\Psi}$ must decay as
$\mathcal{O}(1/(r_1^2+r_2^2))$ at infinity. Thus, we define
\begin{equation}
  \dot{\psi}\equiv \frac{r_1^{2}+r_2^2+\tilde{R}^2}{R^2}\dot{\Psi}\ ,\qquad
 \ddot{\psi}\equiv \frac{r_1^{2}+r_2^2+\tilde{R}^2}{R^2} \ddot{\Psi}\ .
\label{fac2}
\end{equation}
and impose Neumann boundary conditions for $\dot{\psi}$ and
$\ddot{\psi}$ at infinity. Note that, in the above equation, we factorized
$r_1^2+r_2^2+\tilde{R}^2$ instead of $r_1^2+r_2^2$ to make the transformation regular at the
origin $r_1=r_2=0$.

We choose our inner boundary $U$ to be the surface $z/z_\textrm{max}=-0.5$, where we impose the following Dirichlet
boundary conditions:
\begin{equation}
\label{ringpert}
 \dot{\psi}|_U = 1\ ,\quad  \ddot{\psi}|_U = 0\ .
\end{equation}
As we shall see in the next subsection, this simple boundary condition is sufficient to capture the instability of  fat rings.  At the axes $r_1=0$ or $r_2=0$, we impose Neumann boundary conditions as
\begin{equation}
 \partial_{r_1}\dot{\psi}|_{r_1=0}= \partial_{r_1}\ddot{\psi}|_{r_1=0}=
 \partial_{r_2}\dot{\psi}|_{r_2=0}=
 \partial_{r_2}\ddot{\psi}|_{r_2=0}=0\ .
\end{equation}

\begin{figure}[htbp]
  \centering
  \subfigure
  {\includegraphics[height=5.0cm,clip]{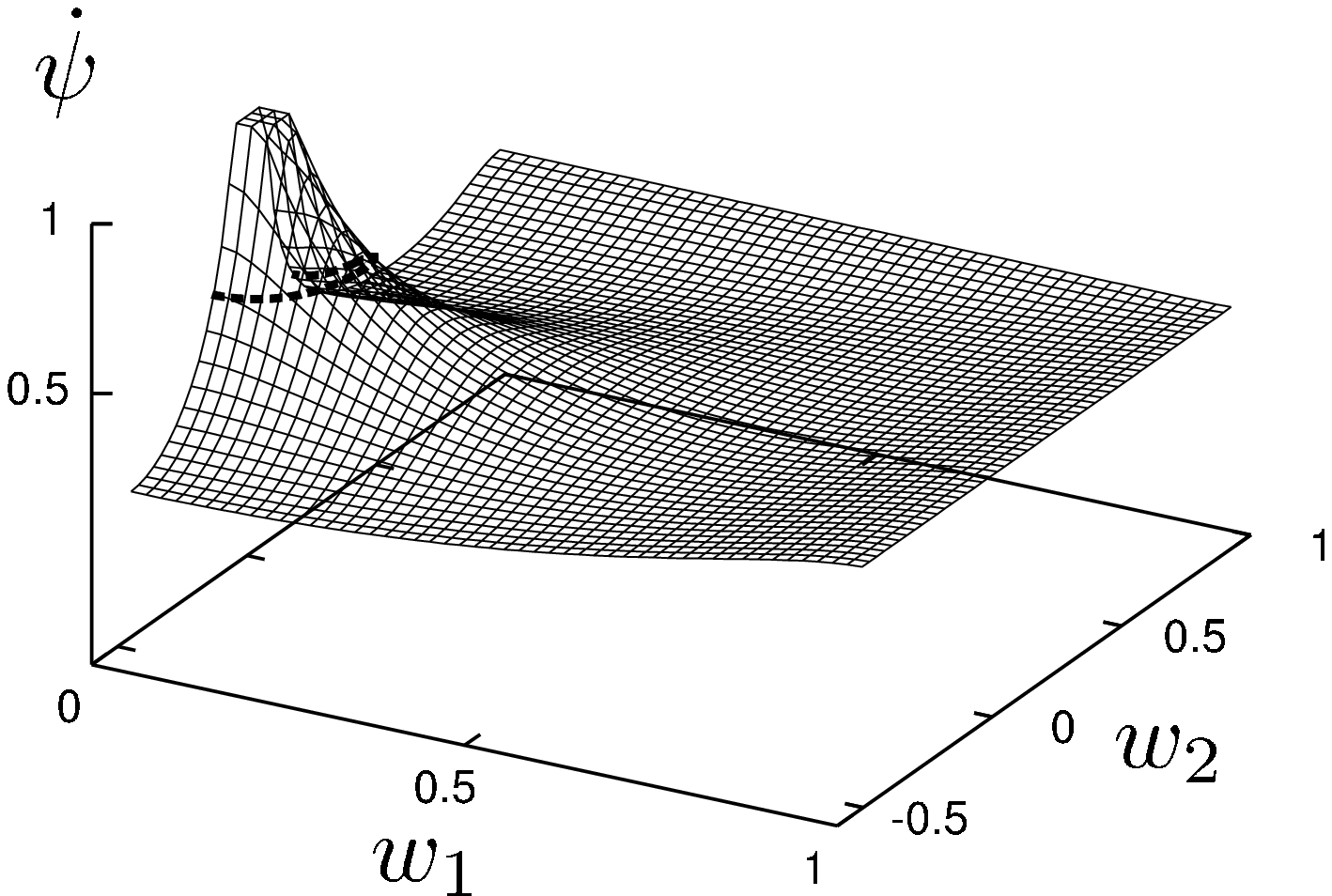}
    \label{fig:dpsiBR}
  }
  \subfigure
  {\includegraphics[height=5.0cm,clip]{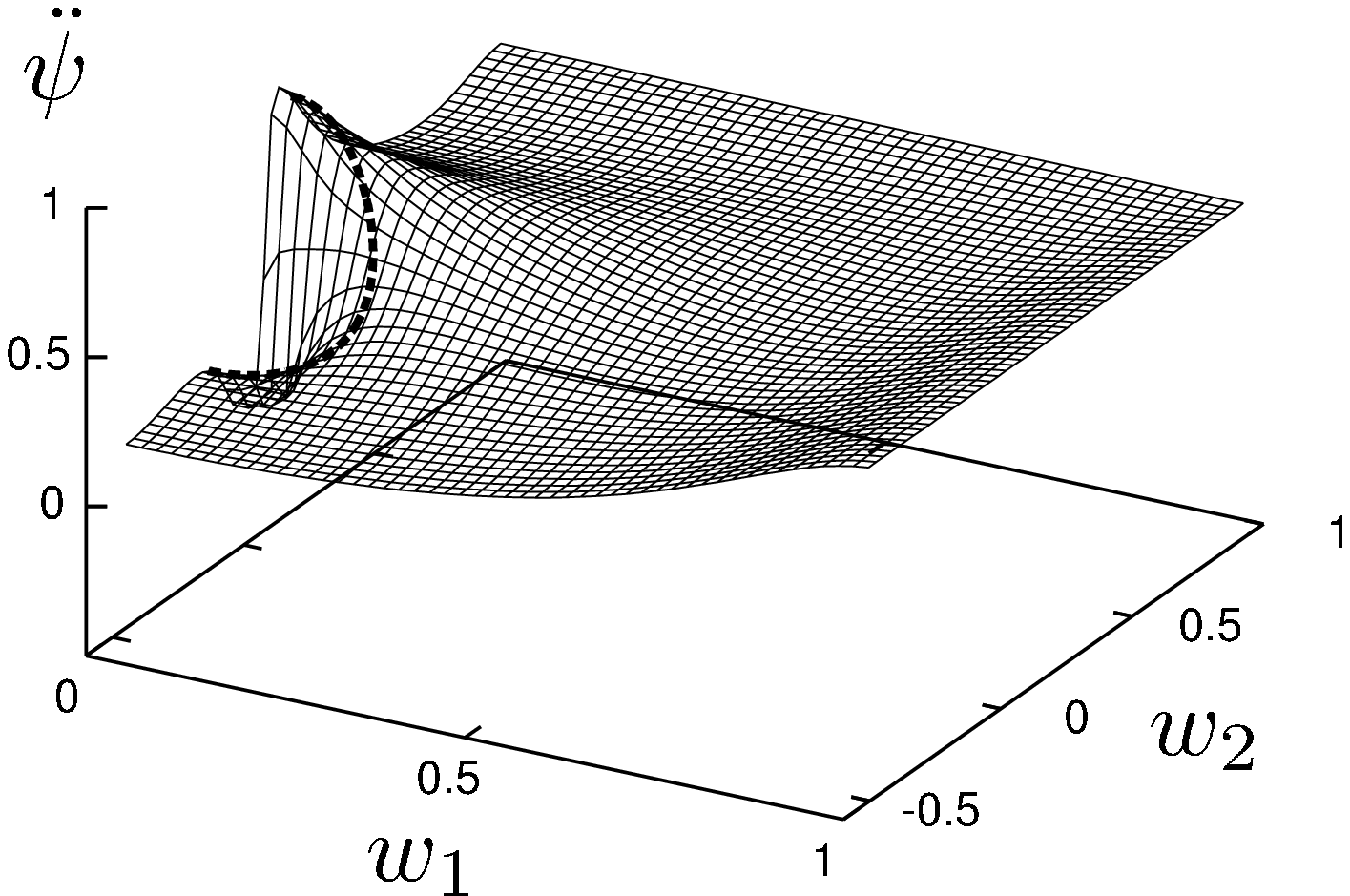} 
    \label{fig:ddpsiBR}
  }
  \caption{\label{fig:psiBR}
The solutions of $\dot{\psi}$ (left) and $\ddot{\psi}$ (right) for
 $\nu=0.6$. The dotted line is the position of the horizon of the unperturbed solution ($z=0$).
}
\end{figure}

Furthermore, in our numerical calculation, we introduce compact coordinates $w_1$ and $w_2$
defined as
\begin{equation}
 r_1=c \tan\left(\frac{\pi}{2}w_1\right)\ ,\quad
 r_2-\tilde{R}=c \tan\left(\frac{\pi}{2}w_2\right)\ .
\label{w1w2}
\end{equation}
We prepare grids at even intervals in the $w_1$ and $w_2$ coordinates.
The constant $c$ is chosen so that $N_\textrm{in}/N_\textrm{all}\simeq 0.025$,
where $N_\textrm{all}$ represents total number of grid points and
$N_\textrm{in}$ is the number of grid points inside the horizon $z<0$.
The coordinate ranges are $0<w_1<1$ and $2\tan^{-1}(-\tilde{R}/c)/\pi<w_2<1$.
Using the $w_1$ and $w_2$ coordinate, we solve Eqs.(\ref{Psi1}) and
(\ref{Psi2}). As an example, we show solutions for $\nu=0.6$ in
Figure \ref{fig:psiBR}. Note that $\dot{\psi}$, $\ddot{\psi}$ and $w_1,w_2$ are dimensionless. It follows that our results depend only on the dimensionless parameter $\nu$, not on the scale $R$ (or, equivalently, the mass $M$). 
 
 From the asymptotic values of $\dot{\psi}$ and $\ddot{\psi}$,
we can determine the deviation of mass $\dot{M}(0)$ and $\ddot{M}(0)$
using Eq.(\ref{dotM}). 

\begin{figure}[htbp]
  \begin{center}
   \includegraphics[height=5.0cm,clip]{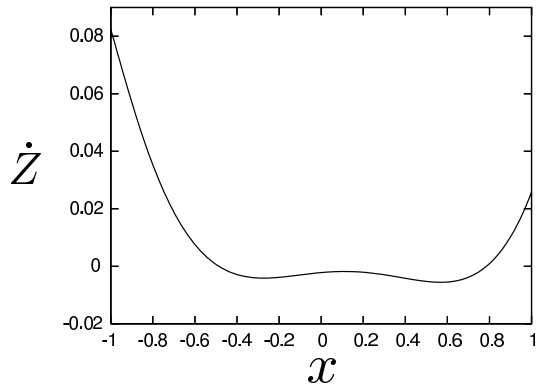}
   \caption{
\label{fig:zahBR}
The first order deviation of the position of the apparent horizon for the solution of Fig. \ref{fig:psiBR}. 
}
  \end{center}
\end{figure}

In the background solution, the apparent horizon is at $z=0$. At first order this will be displaced to some new position
$z=\lambda \dot{Z}(x)$. This is found by solving equation (\ref{deltaEL}) of Appendix \ref{sec:horizonarea}, which takes the same form as equation~(\ref{appeq}), where the 
functions $A$, $B$ and $C$ are given by 
\begin{equation}
\begin{split}
&A=-\frac{2x+\nu(1+x^2)}{(1-x^2)(\nu x+1)}\ ,\\
&B=\frac{1+\nu}{4\lambda(1+\lambda)(1-x^2)(\nu x+1)^3}
\big[
2\lambda \nu x^2
+2\lambda\nu\{
(1+3\lambda)\nu-(1+\lambda)
\}x\\
&\hspace{5cm}
-(1-\lambda)\nu^2
-(1-4\lambda-5\lambda^2)\nu
-3\lambda(1+\lambda)
\big]\ ,\\
&C=-\frac{3(1-\nu^2)}{4\nu(1-x^2)(\nu x+1)}\ .
\end{split}
\label{ABCBR}
\end{equation}
Fig.\ref{fig:zahBR} shows the position of the apparent horizon for the solution $\dot{\psi}$ of Fig. \ref{fig:psiBR}. 

The quantities $\dot{A}_\textrm{app}$ and $\ddot{A}_\textrm{app}$ are calculated as explained in Appendix \ref{sec:horizonarea}. The results are given by eqs (\ref{darea}) and
(\ref{ddarea}) with the functions $I_0$, $I_1$ and $I_2$ now being given by
\begin{equation}
\begin{split}
&I_0(x)=\frac{12\pi^2\nu\sqrt{\nu\lambda(1-\lambda^2)}}{(1-\nu)(1+\nu
 x)^2}R^3\ ,\\
&I_1(x)=\frac{16\pi^2 \nu^3 (1-x^2)R^3}{(1+\nu)(1-\nu)^2(1+\nu
  x)}\sqrt{\frac{\lambda(1-\lambda^2)}{\nu}}\ ,\\
&I_2(x)=\frac{4\pi^2 \nu^3 R^3}{(1-\nu)^2(1+\nu
  x)^4}\sqrt{\frac{1-\lambda}{\nu\lambda(1+\lambda)}}
\big[
2\lambda \nu x^2
+2\lambda\nu\{
(1+3\lambda)\nu-(1+\lambda)
\}x\\
&\hspace{5cm}
-(1-\lambda)\nu^2
-(1-4\lambda-5\lambda^2)\nu
-3\lambda(1+\lambda)
\big]\ .
\end{split}
\label{IBR}
\end{equation}
Substituting our results for $\dot{M}$, $\ddot{M}$, $\dot{A}_\textrm{app}$ and $\ddot{A}_\textrm{app}$
into Eq.(\ref{Qdef}), we can evaluate $Q$.

\subsection{Results}

\begin{figure}[htbp]
  \begin{center}
   \includegraphics[height=6.0cm,clip]{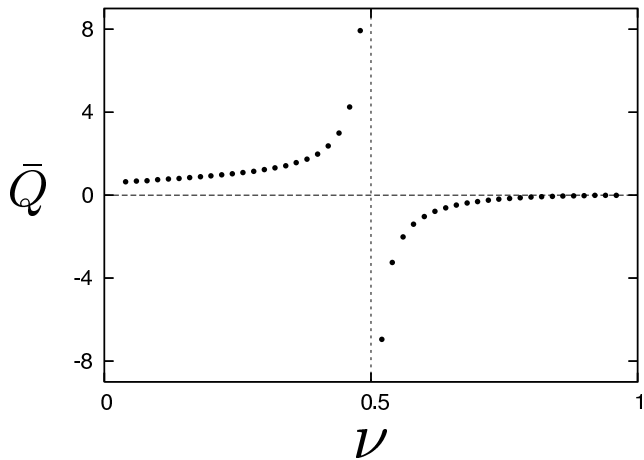}
   \caption{
\label{fig:QBR}
$\bar{Q}$ against $\nu$. $\bar{Q}$ is negative for $\nu>1/2$.
This proves the existence of an instability of fat black rings.
}
  \end{center}
\end{figure}

In Fig.\ref{fig:QBR}, we plot $\bar{Q}$ against $\nu$ where
the quantity $\bar{Q}$ is defined in essentially the same way as for Myers-Perry black holes:
\be
 \bar{Q}\equiv \frac{Q}{D^2 M}, \qquad D^2=\int^1_{-1}\, \frac{dx}{\sqrt{1-x^2}} \dot{\Psi}(z=0,x)^2
\ee
The most striking feature of this plot is the apparent divergence at $\nu=1/2$. This is the property anticipated in the discussion following eq. (\ref{Qdef}): $Q$ diverges because $c_J$ vanishes at $\nu=1/2$. Our results confirm the expectation that fat black rings with $\nu \approx 1/2$ have negative $Q$ and therefore are unstable, as predicted by Ref. \cite{Arcioni:2004ww}. Furthermore, we see that $Q$ is negative at least up to $\nu \approx 0.95$, beyond which the numerical error exceeds the criteria explained in Appendix \ref{sec:Convergence}. Hence all fat rings for which we have reliable results are unstable, in agreement with the prediction of Ref. \cite{Elvang:2006dd}. 

For thin rings, Fig. \ref{fig:QBR} shows that $Q$ is positive so there is no indication of instability. This is for the particular choice of initial data defined by equation (\ref{ringpert}). It might be the case that more complicated initial data leads to $Q<0$. Therefore we have repeated the above calculation for initial data defined by
\begin{equation}
\dot \psi|_U=1+\sum_{n=1}^N\,c_n\,\cos(n\,\theta)\,,
\end{equation}
where $\cos\theta =x$, and choosing the constants $c_n$ by trying to minimize $\bar{Q}$, as we did for the Myers-Perry black holes. We took $N=4$ and considered  values of $\nu$ such that $0.05\leq \nu\leq 0.45$. However, in all cases we found $Q>0$ so there is no indication of any rotationally symmetric instability of thin rings.

\section{Doubly spinning black rings}
\subsection{Background solution}

We can study the instability of doubly spinning black rings 
in the same way as for the singly spinning black rings. The metric of the (balanced) doubly spinning black ring is given
in Ref. \cite{Pomeransky:2006bd}. It is labelled by a constant $k$ with dimensions of length, and two dimensionless constants $(\alpha,\nu)$\footnote{
Here $(\alpha,\nu)$ correspond to $(\nu,\lambda)$ of Ref.~\cite{Pomeransky:2006bd}. The $(\phi,\psi)$ coordinates of this reference correspond to our $(\tilde\phi_1,\tilde\phi_2)$.} which lie in the range
\begin{equation}
 0\leq \alpha < 1\ ,\quad 2\sqrt{\alpha}\leq \nu <1+\alpha\ .
\end{equation}
In the second inequality, the lower limit corresponds to an extreme horizon and the upper limit to a naked singularity.
When $\alpha=0$ the solution reduces to the singly spinning ring. 

The solution is written using coordinates $(t,x,y,\tilde{\phi_1},\tilde{\phi_2})$.
The coordinates $(\tilde{\phi_1},\tilde{\phi_2})$ have canonical periodicity $2\pi$ and the same interpretation as for singly spinning rings (e.g., the $S^1$ of the ring horizon is tangent to $\partial/\partial\tilde{\phi_2}$). The event horizon is located at
\begin{equation}
 y=\frac{-\nu+\sqrt{\nu^2-4\alpha}}{2\alpha} \equiv y_h\ .
\end{equation}
The mass, angular momenta, temperature and angular velocities are given in Ref. \cite{Pomeransky:2006bd}.
We find that the heat capacity at constant angular momenta is
\begin{equation}
\begin{split}
c_J&=
\frac{24k^3\pi^2(1-\alpha y_h) (1+\nu y_h) (1+\alpha y_h^2)^2
\{-2\nu^2-(1+\alpha)\nu+1+10\alpha+\alpha^2)\}
}
{\alpha(1+y_h)(1-\alpha)}
\\
&\times\big[
1
-(1+13\alpha)y_h
+(2+31\alpha+8\alpha^2)y_h^2
+(-2-40\alpha-10\alpha^2+4\alpha^3)y_h^3
\\
&
+2\alpha(2-28\alpha+5\alpha^2)y_h^4
+2\alpha(1+\alpha)(5+21\alpha+5\alpha^2)y_h^5
+2\alpha^2(5-28\alpha+2\alpha^2)y_h^6
\\
&
-2\alpha^2(-2+5\alpha+20\alpha^2+\alpha^3)y_h^7
+\alpha^3(8+31\alpha+2\alpha^2)y_h^8
-\alpha^4 (13+\alpha)y_h^9
+\alpha^5y_h^{10}
\big]^{-1}
\end{split}
\end{equation}
This vanishes at $\nu=\nu_0$ where
\begin{equation}
 \nu=\frac{1}{4}(-1-\alpha+\sqrt{(\alpha+9)(9\alpha+1)})\equiv \nu_0\ .
\end{equation}
$c_J$ is negative for $\nu<\nu_0$ and positive for $\nu>\nu_0$. The Jacobian~(\ref{Jac}) also vanishes at $\nu=\nu_0$\footnote{See footnote \ref{DeltacJ} for an explanation of why this must be the case.} and so thin rings are those with $\nu<\nu_0$ and fat rings those with $\nu>\nu_0$.

As in the singly spinning case, we shall consider a surface $\Sigma$ given by a constant $t$ slice of this spacetime, with induced metric $\bar{h}_{ab}$. We extend through the bifurcation surface at $y=y_h$ on $\Sigma$ using the coordinate transformation $z^2=y-y_h$, The coordinate range of the new coordinate $z$ is taken to be $-z_\textrm{max}<z<z_\textrm{max}$ so that $z<0$ corresponds to a new asymptotically flat region and where
$z_\textrm{max}=\sqrt{-1-y_h}$.

\subsection{Results for doubly spinning black rings}

The numerical calculations for doubly spinning black rings are
similar to those for singly spinning black rings and therefore we will only describe them very briefly. 
The coordinate system we use is same as Eq. (\ref{r1r2def}), but 
 the parameter $\tilde{R}$  for doubly spinning black rings is defined as
$\tilde{R}^2=2k^2(1+\alpha-\nu)/(1-\alpha)$.
For this choice of $\tilde{R}$, the metric becomes explicitly 
asymptotically flat as in Eq. (\ref{BRasym}). Introducing new variables 
$(\dot{\psi},\ddot{\psi})=(r_1^{2}+r_2^2+\tilde{R}^2)k^{-2} (\dot{\Psi},\ddot{\Psi})$
 and compact coordinates~(\ref{w1w2}), we impose the boundary condition (\ref{ringpert}) on the surface $U$ given by $z/z_\textrm{max}=-0.5$, as before.

Then, the first order deviation of the apparent horizon can be
determined from the solution of $\dot{\psi}$.\footnote{
For doubly spinning black rings, we omit the explicit equations
corresponding to Eqs. (\ref{ABCBR}) and (\ref{IBR}) because 
they are too long and unilluminating.}
Once we have determined $(\dot\psi, \ddot\psi)$, we can extract the first and second order perturbations of the physical parameters,  $\dot{M}$, $\ddot{M}$,
$\dot{A}_\textrm{app}$, $\ddot{A}_\textrm{app}$
and $Q$.

\begin{figure}[htbp]
  \begin{center}
   \includegraphics[height=6.0cm,clip]{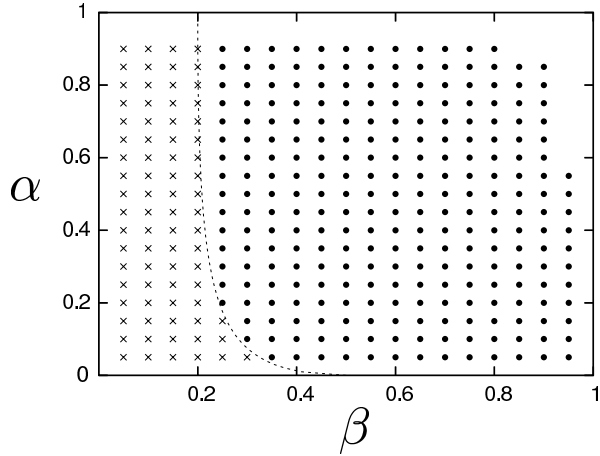}
   \caption{
\label{fig:DSBR}
The signature of $Q$ for various $\alpha$ and $\beta$.
The points $\bullet$ and $\times$ represent $Q<0$ and $Q>0$, respectively.
Points to the left of the dotted curve correspond to thin rings, points to the right correspond to fat rings.}
  \end{center}
\end{figure}

In Fig. \ref{fig:DSBR} we plot the sign of $Q$ (which is what signals the existence of an instability) as a function of parameters $(\nu,\alpha)$ that specify the black ring (as before, the parameter $k$ just sets a scale for $Q$). 
For increased clarity, we have parametrized $\nu$ in terms of a new parameter $\beta$ defined as
\begin{equation}
 \nu=(1+\alpha)\beta+2\sqrt{\alpha}(1-\beta)
\end{equation}
where
$1+\alpha$ and $2\sqrt{\alpha}$ are maximum and minimum values of $\nu$
for a fixed $\alpha$ so  $\beta$ lies in the range $0\le \beta<1$.
The points $\bullet$ and $\times$ correspond to $Q<0$ and $Q>0$, respectively.
In this plot we have included only data which satisfies the numerical accuracy criteria discussed in Appendix \ref{sec:Convergence}. The dotted curve corresponds to $\Delta=0$ (or $c_J=0$). Hence the left and right hand sides of the dotted curve correspond to thin
and fat rings respectively. We can see that fat doubly spinning black rings are unstable. As in the singly spinning case, this is to be expected near to the dotted curve for the reasons explained below (\ref{Qdef}). But our plot also demonstrates that all fat rings (at least those for which we have reliable data) are unstable. There is no sign of any rotationally symmetric instability of thin black rings.

\subsection{Global Penrose inequality and minimally spinning rings}

So far, we have been discussing local Penrose inequalities, i.e., inequalities restricted to initial data describing a small perturbation of a given black hole. However, one can also obtain a {\it global} inequality for black rings. Consider asymptotically flat initial data that, when evolved, "settles down" to a black ring solution with mass $M_F$ and parameters $(\nu_F,\alpha_F)$. Then the usual arguments give
\be
 A_{\rm min} \le A_{\rm ring}(M_F,\nu_F,\alpha_F) \le A_{\rm ring}(E,\nu_F,\alpha_F) 
\ee
where $E$ is the ADM energy of the initial data. The second inequality holds because $A_{\rm ring}$ is an increasing function of mass at fixed $\nu,\alpha$. Now the black ring with greatest horizon area for given mass is the one with $\nu=1/2$, $\alpha=0$ and hence $A_{\rm ring}(E,\nu_F,\alpha_F) \le A_{\rm ring}(E,1/2,0)=(3/16)\sqrt{3/\pi} E^{3/2}$. Hence the inequality
\be
 A_{\rm min} \le a_* E^{3/2}, \qquad a_* =  \frac{3}{16} \sqrt{\frac{3}{\pi}} \approx 0.183
\ee
must apply to any initial data which settles down to a black ring spacetime.\footnote{
This is much more restrictive than the "standard" Penrose inequality, in which the RHS is given by the area of a 5d Schwarzschild black hole of mass $E$, which has the same form with $a_*=(16/3) \sqrt{8\pi/3} \approx 15.4$.}

Initial data describing a ring with $\nu=1/2$ and $\alpha=0$ saturates the above inequality. Such a ring is singly spinning and has $\Delta=0$: it is neither thin nor fat but sits between these two branches. It is "minimally spinning" in the sense that it is the singly spinning ring with smallest angular momentum for a given mass. The stability of a minimally spinning ring is not covered by the analysis of the previous sections (since it is neither thin nor fat). Refs. \cite{essay,Elvang:2006dd} argued that such a ring must be unstable by considering the effect of dropping a particle with zero angular momentum into the ring. This argument neglects the backreaction of the particle. A more rigorous argument for instability can be obtained from the local version of the above Penrose inequality.

Consider a 1-paramter family of asymptotically flat, initial data, with parameter $\lambda$, such that for $\lambda=0$ it reduces to a constant $t$ slice of a minimally spinning ring with mass $M$, angular momentum $J$, angular velocity $\Omega$ and temperature $T$. If the ring is stable then this initial data must satisfy the above inequality for sufficiently small $\lambda$, i.e., 
\be
 a(\lambda)  \equiv \frac{A_{\rm min}(\lambda)}{E(\lambda)^{3/2}} \le  a_*
\ee
for small enough $\lambda$. This is saturated at $\lambda=0$: $a(0)=a_*$. The first correction arises from $\dot{a}(0)$
\bea
 \dot{a}(0) &=& \frac{\dot{A}_{\rm min}(0)}{M^{3/2}}- \frac{3 A_{\rm min}(0)}{2 M^{5/2}} \dot{E}(0) \nonumber \\
 &=& \frac{4}{T M^{3/2}}  \left( \dot{E}(0)-\Omega \dot{J}(0) \right)- \frac{3 \dot{E}(0)}{2 M^{5/2}} \frac{4}{T} \left( \frac{2}{3} M - \Omega J \right) \nonumber \\
&=& \frac {\Omega J}{M^{3/2} T} \left( 6\frac{\dot{E}(0)}{M} - 4 \frac{\dot{J}(0)}{J} \right)
\eea
In the first line we have used $E(0)=M$ and in the second line we have used the first law and the Smarr relation \cite{Emparan:2001wn}. We have checked that there exist perturbations for which $\dot{a}(0) \ne 0$: the methods of the previous sections can be used to construct a perturbation with $\dot{J}(0)=0$, $\dot{E}(0) \ne 0$. If we choose $\lambda$ to have the same sign as $\dot{a}(0)$ then $a(\lambda) = a_* + \lambda \dot{a}(0) + \ldots$ will violate the Penrose inequality for arbitrarily small $\lambda$. Hence minimally spinning rings are unstable.

Note that we were able to reach this conclusion using first order perturbation theory whereas previously we have always had to work to second order. This is related to a breakdown in our observation at the end of section \ref{sec:string} that one can decompose a linearized metric perturbation into non-dynamical and dynamical parts with the former arising from a variation of parameters in the unperturbed black hole, and the latter having vanishing mass and angular momentum at first order. For a minimally spinning ring, it is not possible to do this because a variation of parameters always gives a perturbation with $\dot{a}=0$, i.e., one cannot vary $E$ and $J$ independently this way. The best one can do is to split a perturbation into a non-dynamical part and a dynamical part, such that the latter has, at first order, vanishing energy but, in general, non-vanishing angular momentum. The fact that the dynamical part (which includes any instability) has non-vanishing angular momentum at first order explains why the Penrose inequality is non-trivial at first order.

We have shown that fat rings are unstable against rotationally symmetric perturbations. Therefore fat rings will admit rotationally symmetric linearized perturbations with exponential time-dependence. Such modes should be analytic in $\nu$ (since the background is) and presumably correspond to quasinormal modes of thin rings (since these appear to be stable against rotationally symmetric perturbations). Hence at $\nu=1/2$, by continuity we cannot have an instability which grows exponentially with time: it must be sub-exponential. This is supported by the results of Ref. \cite{Elvang:2006dd}, where the minimally spinning ring corresponds to a point of inflection of the effective potential for radial perturbations. This suggests that the linearized instability should grow linearly with time.

\section{Discussion}

We have described how violation of a local Penrose inequality can be used to demonstrate the existence of certain types of black hole instability. In all cases, we constructed initial data by the Lichnerowicz conformal method. The simplicity of this approach is its main advantage. We did not have to derive and solve a large set of equations governing gravitational perturbations, or worry about gauge issues. Instead one just solves the single linear PDE (\ref{Psi1}) to determine the first order perturbation, and (\ref{Psi2}) to determine the second order perturbation. 

A disadvantage of our approach is that in general it cannot be used to identify precisely the onset of instability. In the black string case, our results were surprisingly accurate. However, in the Myers-Perry example, we found that the local Penrose inequality is violated at a value of the spin parameter somewhat higher than that at which the ultraspinning instability is known to appear. This could be improved by considering more general initial data. 

We have seen how a local Penrose inequality can be used to make the turning point argument for instability (cf Ref. \cite{Arcioni:2004ww}) more rigorous. In this case, the approach does predict precisely when an instability should appear.

The approach we have used here could be used in many other situations. The stability of non-uniform black strings \cite{BS}, or localized Kaluza-Klein black holes \cite{KKBHs}  could be studied this way. One could also study Myers-Perry black holes with multiple non-zero angular momenta.

We have considered only vacuum black holes. The same kinds of argument could be applied to charged black holes. In order to derive a local Penrose inequality, one must have a situation in which charge cannot leave the spacetime (and so the charge of the final state is the same as that of the initial data). For an asymptotically (locally) flat spacetime, this will be the case if there are no charged fields, or all charged fields are massive. The Maxwell field (or, more generally, $p$-form fields) must satisfy the Gauss law constraint. If initial data is constructed by the Lichnerowicz conformal method used above then it is straightforward to solve the Gauss law constraint by suitable conformal rescaling of the electric field of the initial data. 

Our methods are potentially more powerful for asymptotically anti-de Sitter spacetimes. In this case, the usual boundary conditions imply that angular momentum (as well as energy and charge) is conserved. Therefore one would not need to assume rotational symmetry to obtain a local Penrose inequality for rotating black holes. 

Instabilities of charged anti-de Sitter black holes have been the subject of much interest recently \cite{Hartnoll:2011fn}. It would be interesting to consider the various kinds of instabilities that can afflict such black holes using the methods discussed here.

Finally, it would be interesting to relate the approach described here to the approach of linearized stability analysis. Can one prove that violation of the local Penrose inequality implies the existence of a linearized perturbation that grows exponentially with time?

\subsection*{Acknowledgments}

This work was presented at the workshops ``Numerical relativity beyond astrophysics" (ICMS, Edinburgh, UK) and ``Gravity - New perspectives from strings and higher dimensions" (Benasque Centre for Sciences, Spain). PF and HSR would like to thank the organisers and the participants of these workshops for discussions and comments. KM is supported by a grant for research abroad by the JSPS (Japan). PF is supported by an EPSRC postdoctoral fellowship [EP/H027106/1].  HSR is a Royal Society University Research Fellow.

\appendix

\section{Apparent horizon for $t-\phi_i$ symmetric initial data}

\label{sec:apparent}

We will show that, for $t-\phi_i$ symmetric initial data, the apparent horizon $S$ is a local minimum of area in the set of homologous surfaces outside $S$ which are tangent to the Killing fields $\Phi^{ia}$. 

{\it Proof}. We have shown that the apparent horizon is a minimal surface, i.e., extremizes the area. So we need to consider the second variation of the area. Let $\phi$ be a positive scalar on $S$ and define the vector field $V^a$ on $S$ by $V^a = \phi n^a$ where $n^a$ is the outward unit normal to $S$ in $\Sigma$. The map $p \rightarrow \exp(tV)$ then defines a 1-parameter deformation of $S$ specified by the function $\phi$. Since $S$ is minimal, the first derivative of the area vanishes under this deformation. The second derivative is
\be
\label{d2A}
\left(  \frac{d^2 A}{dt^2} \right)_{t=0} = \int_S \phi \left( \frac{\partial \theta}{\partial t} \right)_{t=0}
\ee 
where $\theta$ denotes the expansion of the outward null normal to the deformed surface. Ref. \cite{Cai:2001su} showed that
\be
\left( \frac{\partial \theta}{\partial t} \right)_{t=0} = - \triangle \phi + 2 X^a D_a \phi + \left(Q + D_a X^a - X_a X^a \right) \phi \equiv {\cal O}(\phi)
\ee
where $\triangle$ denotes the Laplacian on $S$, $D_a$ is the metric connection on $S$, indices are raised and lowered on $S$,  $Q$ is a certain scalar on $S$ depending on the (instrinsic and extrinsic) curvature of $S$ and $X^a$ is the vector field
\be
 X^a = (h^{ab} - n^a n^b) K_{bc} n^c.
\ee
In the time-symmetric case, $X^a=0$ and ${\cal O}$ is self-adjoint with respect to the obvious inner product. In this case one can argue that $S$ must be a stable minimal surface, i.e., (\ref{d2A}) is non-negative \cite{Cai:2001su}. If $X^a \ne 0$ then ${\cal O}$ is not self-adjoint but its eigenvalues have real part bounded below, the eigenvalue $\lambda_1$ with the smallest real part is called the principal eigenvalue and must be real and non-negative \cite{Andersson:2005gq,Galloway:2005mf}. For the initial data considered in section \ref{rotating}, $X^a$ is a linear combination of the Killing vectors
\be
 X^a = n^b J^i_b \Phi^{ia}
\ee
We shall consider only deformations of the surface for which the Killing vectors $\Phi^i_a$ remain tangent to the surface. This is equivalent to considering only functions $\phi$ invariant with respect to these Killing fields. Hence we have $X^a D_a \phi=0$. Restricted to the space of such functions, ${\cal O}$ is self-adjoint. Therefore the eigenvalues of the restriction are real. Since they cannot be less than $\lambda_1$, they must be non-negative. Expanding $\phi$ in (\ref{d2A}) in eigenfunctions we deduce that $d^2 A/dt^2 \ge 0$.

Of course we really would like to establish the strict inequality $d^2 A/dt^2>0$. This could be done if we knew $\lambda_1 \ne 0$. To prove this we could show $\lambda_1 \ne 0$ for the unperturbed spacetime, then it will also hold in the perturbed spacetime. Or if $\lambda_1=0$ in the unperturbed spacetime we would expect the perturbed spacetime to satisfy $\lambda_1 \ne 0$. Either way it is clear that $S$ should be a local minimum of area.

\section{Apparent horizon area}

\label{sec:horizonarea}

In this Appendix, we will explain how to determine the second order change in the area of the apparent horizon. Although many of the formulae here could be written in a covariant form, we will derive them in a coordinate system adapted to the symmetries of the black holes of interest. 

Consider $(d-1)$-dimensional initial data which is cohomogeneity-2, i.e., it depends non-trivially on at most two coordinates. We shall denote these coordinates as $(r,x)$ and assume $a \le x \le b$. We assume that the initial data has a metric of the form
\be
\label{conftrans}
 h_{ab} = \Omega(\lambda,r,x)^2 \bar{h}_{ab}
\ee
where $\bar{h}_{ab}$ is initial data for the black hole whose stability we are investigating. Furthermore we assume $\Omega(0,r,x)=1$, so $h_{ab}=\bar{h}_{ab}$ when $\lambda=0$. In the cases we are interested in, $r$ will be a "radial" coordinate. For the black string, $x$ will be the coordinate around the KK circle and for the rotating black holes, $x$ will be a direction cosine. Note that $h_{ab}$ depends on $\lambda$ only through the conformal factor $\Omega$.

Now consider a surface $r=r(x)$ in this initial data. Its area is given by a functional of the form (using a prime to denote a $x$-derivative)
\be
 {\cal A}[\lambda,r(x),r'(x)] = \int_a^b dx \, F(\lambda,x,r(x),r'(x)).
\ee
Note that (\ref{conftrans}) implies that
\be
\label{FbarF}
 F(\lambda,x,r(x),r'(x)) = \Omega(\lambda,x,r(x))^{d-2} \bar{F}(x,r(x),r'(x))
\ee
where $\bar{F}(x,r(x),r'(x)) = F(0,x,r(x),r'(x))$. In the cases we are interested in, the apparent horizon is a minimal surface, i.e., it extremizes ${\cal A}$. Let $r=R(\lambda,x)$ denote this minimal surface, which must satisfy the Euler-Lagrange equation
\be
\label{Eeq}
 E(\lambda,x,R(\lambda,x),R'(\lambda,x),R''(\lambda,x)) = 0,
\ee
where
\be
 E(\lambda,x,r(x),r'(x),r''(x)) \equiv \frac{d}{dx} \frac{\partial F}{\partial r'} - \frac{\partial F}{\partial r}.
\ee
The area of the apparent horizon is 
\be
 A_{\rm app}(\lambda) = {\cal A}[\lambda,R(\lambda,x),R'(\lambda,x)].
\ee
Using the Euler-Lagrange equation, we have
\be
 \dot{A}_{\rm app}(\lambda) = \int_a^b dx \dot{F}(\lambda,x,R(\lambda,x),R'(\lambda,x))+ \left[ \left( \frac{\partial F}{\partial r'} \right)_{r(x)=R(\lambda,x)} \dot R(\lambda,x) \right]_a^b
\ee
where we are using a dot to denote $\partial/\partial \lambda$ (which acts just on the first argument of $F$). The second term is a surface term coming from the boundaries of the $x$-integration. For the black string, we impose periodic boundary conditions so the surface term vanishes. For the rotating black holes, we use $\partial F/\partial r' = \Omega^{d-2} \partial {\bar F}/{\partial r'}$. Functions $r(x)$ describing a smooth surface will satisfy $ \partial {\bar F}/{\partial r'} =0$ at $x=a,b$. Hence we can neglect the surface term.

Taking another derivative gives
\be
\label{ddotA1}
 \ddot{A}_{\rm app}(\lambda) = \int_a^b dx \left[ \ddot{F} (\lambda,x,R(\lambda,x),R'(\lambda,x)) - \left( \frac{d}{dx} \frac{\partial \dot{F}}{\partial r'} - \frac{\partial \dot{F}}{\partial r} \right)_{r(x)=R(\lambda,x)} \dot{R}(\lambda,x) \right]
\ee
where $\bar{R}(x) = R(0,x)$. A surface term can be neglected for the same reason as before.
To simplify this expression, note that (\ref{FbarF}) implies
\be
\label{Flambda}
 \dot{F}(\lambda,x,r(x),r'(x)) = (d-2) \frac{\dot{\Omega}(\lambda,x,r(x))}{\Omega(\lambda,x,r(x))} F (\lambda,x,r(x),r'(x))
\ee
and
\be
 \ddot{F}(\lambda,x,r(x),r'(x)) = (d-2) \left( \ddot{\Omega}(\lambda,x,r(x)) +(d-3) \dot{\Omega}(\lambda,x,r(x))^2 \right) \bar{F}(x,r(x),r'(x))
\ee
The first of these implies
\be
\label{Adotzero}
 \dot{A}_{\rm app}(0) = (d-2) \int_a^b dx \, \dot{\Omega}(0,x,\bar{R}(x)) \bar{F}(x,\bar{R}(x),\bar{R}'(x))
\ee
To evaluate the second term in (\ref{ddotA1}), we use (\ref{Flambda}) and the equation of motion (\ref{Eeq}) to obtain
\bea
  \left( \frac{d}{dx} \frac{\partial \dot{F}}{\partial r'} - \frac{\partial \dot{F}}{\partial r} \right)_{r(x)=R(\lambda,x)} &=& (d-2) \left[ \frac{d}{dx} \left( \frac{\dot{\Omega}}{\Omega} \right) \frac{\partial F}{\partial r'} - \frac{\partial}{\partial r}   \left( \frac{\dot{\Omega}}{\Omega} \right)  F \right]_{r(x)=R(\lambda,x)}  \nonumber \\
&=& (d-2) \left[  \frac{\partial}{\partial r}   \left( \frac{\dot{\Omega}}{\Omega} \right) \left( r' \frac{\partial F}{\partial r'} - F \right) +  \frac{\partial}{\partial x} \left( \frac{\dot{\Omega}}{\Omega} \right) \frac{\partial F}{\partial r'} \right]_{r(x)=R(\lambda,x)} 
\eea
Substituting into (\ref{ddotA1}) and evaluating at $\lambda=0$ now gives
\bea
\label{ddotA}
 \ddot{A}_{\rm app}(0)&=&(d-2) \int_a^b dx \left \{ \left[\ddot{ \Omega} +(d-3) \dot{\Omega}^2 \right]_{\lambda=0,r(x)=\bar{R}(x)} \bar{F}(x,\bar{R}(x),\bar{R}'(x)) \right. \nonumber \\ 
 &-&  \left. \left[  \left(\frac{\partial \dot{\Omega}}{\partial r} \right)_{\lambda=0} \left( r' \frac{\partial \bar{F}}{\partial r'} - \bar{F} \right) + \left( \frac{\partial \dot{\Omega}}{\partial x}  \right)_{\lambda=0} \frac{\partial \bar{F}}{\partial r'} \right]_{r(x)=\bar{R}(x)} \dot{R}(0,x)
   \right\}
\eea
where $\bar{R}(x) = R(0,x)$. Note that this expression involves the metric perturbation to second order in $\lambda$ (via $\ddot{\Omega}$) but we only need to determine the {\it first} order perturbation $\dot{R}(0,x)$ to the position of the apparent horizon. 

Next we need to determine $\dot{R}(0,x)$. To do this, we substitute (\ref{FbarF}) into the minimal surface equation (\ref{Eeq}). The result is
\be
\left[ \Omega^{d-2} \bar{E}(x,r(x),r'(x),r''(x) ) + \frac{\partial}{\partial x} \left( \Omega^{d-2} \right) \frac{\partial \bar{F}}{\partial r'} + \frac{\partial}{\partial r} \left( \Omega^{d-2} \right) \left( r'\frac{\partial \bar{F}}{\partial r'}  - \bar{F} \right) \right]_{r=R(\lambda,x)}= 0
\ee
where $\bar{E}(x,r(x),r'(x)) = E(0,x,r(x),r'(x),r''(x))$. Evaluating at ${\cal O}(\lambda)$ gives
\bea
\label{deltaEL}
&{}&\left[ \frac{d}{d\lambda} \bar{E}(x,R(\lambda,x),R'(\lambda,x),R''(\lambda,x))  \right]_{\lambda=0} \nonumber \\ &{}&+ (d-2) \left[\left(  \frac{\partial \dot{\Omega}}{\partial x} \right)_{\lambda=0} \frac{\partial \bar{F}}{\partial r'} + \left( \frac{\partial \dot{\Omega}}{\partial r}\right)_{\lambda=0} \left( r' \frac{\partial \bar{F}}{\partial r'} - \bar{F}\right) \right]_{r=\bar{R}(x)}=0
\eea
The first term gives an expression linear in $\dot{R}(0,x)$ and its first and second $x$-derivatives. The second term is a source depending on the first order metric perturbation $\dot{\Omega}$. Hence this equation is a second order linear ODE that, with suitable boundary conditions, determines $\dot{R}(0,x)$. 

We now apply this to the $d=n+3$ dimensional  black string \eqref{eqn:SchwBS}, for which 
\be
 \bar{F}(x,r(x),r'(x)) = \frac{\omega_n\,r_+^n}{(1-y(x)^2)^n}  \sqrt{1+\frac{4\,r_+^2(y'(x))^2}{g(y(x))(1-y(x)^2)^4}}
\ee
where $\omega_n$ denotes the area of a unit round $n$-sphere,  $\Omega=\Psi^{\frac{2}{n}}$  and $\Psi$ is given by equation (\ref{Psisol}). The horizon is at $y=0$ in the unperturbed spacetime hence $\bar{R}(x) = 0$. Equation (\ref{Adotzero}) gives 
\be
\dot{A}_{\rm app} (0) \propto \int_0^{2\pi L}  dx \cos (x/L) = 0
\ee
so the first order change in $A_{\rm app}$ vanishes, as discussed above. Equation (\ref{deltaEL}) reduces to
\be
 \left[   \dot{R}''(0,x) - \frac{n(n-1)}{2\, r_+^2} \dot{R}(0,x) \right] = \frac{n^2-1}{2\,n\,r_+^2} \left( \frac{df}{dy} \right)_{y=0} \cos(x/L) 
\ee
The general solution of this equation is a sum of a particular integral proportional to $\cos(x/L)$ and an arbitrary linear combination of $e^{\pm x\sqrt{n(n-1)}/(r_+\sqrt 2)}$. Since no such linear combination obeys the required periodicity $x \sim x+2\pi L$, the only acceptable solution is
\be
 \dot{R}(0,x) = - \frac{n^2-1}{n[n(n-1)+2\,r_+^2/L^2]} \left( \frac{df}{dy} \right)_{y=0} \cos(x/L)\,. 
\ee
Hence this is the first order perturbation to the position of the apparent horizon. We now substitute this into (\ref{ddotA}) to obtain the second order change in the apparent horizon area (\ref{d2Astring}).

\section{Numerical errors}
\label{sec:Convergence}

In this appendix we present some of the convergence tests that we have performed in order to check our numerics. As discussed in the text, the main quantity that we have monitored to check numerical errors is the First Law, which we know  must be satisfied in the continuum limit and therefore it provides a global measure of the numerical error. 

The mass of the background spacetime $M(0)$ sets the scale of problem, but the actual ``size" of the perturbation can be captured by the quantity
\begin{equation}
D^2=\int_{H}\dot \Psi|_H^2\,,
\end{equation}
which is evaluated on the horizon of the unperturbed spacetime. In order to measure the numerical error we want to consider a dimensionless quantity for which the size of the perturbation has been scaled out. Therefore, the following quantity 
\begin{equation}
\epsilon=\frac{1}{D\,M(0)}\bigg|\dot M(0)-\frac{T\,\dot A_\textrm{app}(0)}{4}\bigg|\,,
\end{equation}
should be a good measure of the numerical error and this is what we have used. We reject data for which $\epsilon > 0.01$. Furthermore, $\epsilon$ is an estimate of the error in $\bar{Q}$ because the equation for $\ddot{\Psi}$ is the same as that for $\dot{\Psi}$ but with a source. Since we want to know whether $\bar{Q}$ is positive or negative, we reject data for which $|\bar{Q}|<100 \epsilon$ (this is only an issue for black rings, e.g. $|\bar{Q}|$ becomes small as $\nu \rightarrow 0$ or $1$ for singly spinning rings).

\begin{figure}[t]
\begin{center}
\includegraphics[scale=1]{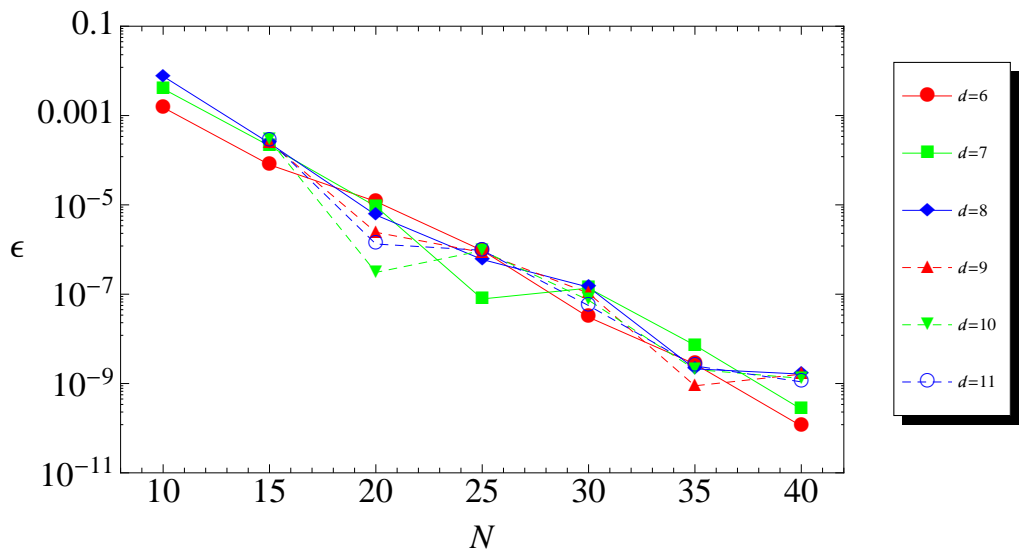}
\end{center}
\caption{Convergence tests for  MP black holes in all dimensions. The error decreases exponentially as a function of the number of grid points independently of the number of spacetime dimensions $d$. }
\label{fig:convergenceMP}
\end{figure}

We have written two codes, one based on second order finite differencing approximation and the other based on a pseudospectral collocation approximation. In Figure \ref{fig:convergenceMP} we present the results of the convergence tests that we have performed for the pseudospectral code for  Myers-Perry black holes in all dimensions.\footnote{For singly spinning rings and doubly spinning rings the results are qualitatively similar and we do not show them here.} As this figure shows, the error $\epsilon$ defined above decreases exponentially with the number of grid points $N$, as expected for smooth functions. It is interesting to note that for $d\geq 9$ our code could not find acceptable solutions for low resolutions, i.e., 10 grid points, but for higher resolutions the behaviour of our code seems to be independent of the number of spacetime dimensions. To carry out the tests we have chosen the number of grid points to be the same in  all coordinate directions. Finally we note that we have also observed second order convergence for the finite differencing code. Therefore, we conclude that our codes exhibit the expected convergence to the continuum according to each differentiation scheme and hence we believe that the numerics are under control.   

\end{document}